\documentclass[useAMS,usenatbib]{mn2e}

\usepackage{graphicx,subfigure}
\usepackage{multirow}
\usepackage{amssymb}

\begin{document}

\title[Synthetic CO, H$_{2}$ and H{\sc{i}} surveys of the Galactic 2nd Quadrant]
{Synthetic CO, H$_{2}$ and H{\sc{i}} surveys of the Galactic 2nd Quadrant, and the properties of molecular gas}

\author[A. Duarte-Cabral et al.]{A. Duarte-Cabral$^1$\thanks{E-mail: adc@astro.ex.ac.uk}, D. M. Acreman$^1$, C. L. Dobbs$^1$, J. C. Mottram$^2$, S. J. Gibson$^3$,  \newauthor C. M. Brunt$^1$, and K. A. Douglas$^4$ \\
$^1$ School of Physics and Astronomy, University of Exeter, Stocker Road, Exeter, EX4 4QL, U.K. \\
$^2$ Leiden Observatory, Leiden University, PO Box 9513, 2300 RA Leiden, The Netherlands \\
$^3$ Department of Physics and Astronomy, Western Kentucky University, 1906 College Heights Blvd., Bowling Green, KY 42101, USA \\
$^4$ Okanagan College, Department of Physics and Astronomy, 1000 KLO Road, Kelowna, British Columbia, Canada
}

\maketitle

\begin{abstract}

We present CO, H$_{2}$, H{\sc{i}} and HISA distributions from a set of simulations of grand design spirals including stellar feedback, self gravity, heating and cooling. We replicate the emission of the 2nd Galactic Quadrant by placing the observer inside the modelled galaxies and post process the simulations using a radiative transfer code, so as to create synthetic observations. We compare the synthetic datacubes to observations of the 2nd Quadrant of the Milky Way to test the ability of the current models to reproduce the basic chemistry of the Galactic ISM, as well as to test how sensitive such galaxy models are to different recipes of chemistry and/or feedback. 

We find that models which include feedback and self-gravity can reproduce the production of CO with respect to H$_{2}$ as observed in our Galaxy, as well as the distribution of the material perpendicular to the Galactic plane. While changes in the chemistry/feedback recipes do not have a huge impact on the statistical properties of the chemistry in the simulated galaxies, we find that the inclusion of both feedback and self-gravity are crucial ingredients, as our test without feedback failed to reproduce all of the observables. Finally, even though the transition from H$_{2}$ to CO seems to be robust, we find that all models seem to underproduce molecular gas, and have a lower molecular to atomic gas fraction than is observed. Nevertheless, our fiducial model with feedback and self-gravity has shown to be robust in reproducing the statistical properties of the basic molecular gas components of the ISM in our Galaxy.

\end{abstract}


\section{Introduction}

Observations of the wide-scale interstellar medium (ISM) have seen tremendous improvements in recent years, with the inception of large surveys to trace the global distribution of the gas and dust between stars in our Galaxy. The bulk of the Galactic structure is delineated through observations of the 21-cm H{\sc{i}} transition, which traces the neutral atomic gas quite reliably. Surveys such as those comprising the International Galactic Plane Survey \citep{Taylor03,McCG2005,Stil06} and more recent all-sky efforts such as EBHIS \citep{kerp2011}, GASS \citep{McCG2009} and GALFA-HI \citep{Peek2011} have surpassed prior H{\sc{i}} surveys in terms of resolution and sensitivity.  The cold molecular gas, which is chiefly H$_2$, remains indirectly and imperfectly traced, primarily with the $J=1-0$ transition of CO gas \citep[e.g.][]{Dame01}, or by using dust as a proxy for total hydrogen column density $N$(H$_{\rm total}$), and comparing that measure to $N$(H), the column density of the atomic component. 

In the absence of a direct measure of global $N$(H$_2$) one must always ask how much molecular gas is being missed by the use of CO and/or dust as a proxy of H$_2$.  Conditions where CO may be absent from the ISM in regions replete with H$_2$ are of two general sorts: (i) photon-dominated regions where CO is dissociated while H$_2$ is more successful in self-shielding, and (ii) cold (T $\lesssim$ 25\,K), well-shielded dense regions where CO is frozen onto dust grains, depleting it from the gas phase.

An additional tracer of potential star forming material is H{\sc{i}} self-absorption (HISA) which traces cold atomic hydrogen \citep{Gibson2000,Gibson10}, a precursor to the transition to the molecular phase. However the relationship between HISA, CO emission and H$_2$ is complex, as HISA is geometry dependent requiring that cold material is viewed against a bright background with the same line of sight velocity.

Simulations have recently started including H$_{2}$ formation and being analysed using observed tracers \citep[e.g.,][]{douglas_2010,Shetty2011a,Acreman12,Smith2014,Kim2014,Pettitt2014}. 
Unlike observations, simulations provide exact measurements rather than estimates (e.g. of the conversion factor between CO intensities and H$_{2}$ column densities, X$_{\rm CO}$), and simulations do not suffer from ambiguities related to the use of kinematic distances. There are now many simulations of isolated galaxies investigating the ISM, but to truly compare with observations they ideally need to be converted to observed tracers such as H{\sc{i}} and CO. This requires a radiative transfer code to produce data cubes of the emission. Here we present an analysis of galactic scale CO, H$_{2}$ and H{\sc{i}} distributions, by using a radiative transfer code to post process simulations of grand design spirals including stellar feedback, self gravity, heating and cooling.

For this paper we use a set of simulated galaxies (which include a chemistry model) to generate synthetic Galactic plane surveys in the $^{12}$CO$(1-0)$ transition and the 21~cm H{\sc{i}} line, and compare those to real observations of the Milky Way. For this comparison, we chose to focus on the second quadrant of the Galaxy, both due to its less complex structure (compared to the inner Galaxy), and the availability of the relevant observational datasets. This allows us to test the ability of the current models in reproducing the basic chemistry of the Galactic ISM, and retrieve the best chemical representation of the Galaxy. We do so by determining the relationship between $\rm{H}_2$, CO emission and {H\sc{i}} self-absorption. We also test how sensitive such galaxy models are to different recipes of chemistry and/or feedback. In Section~\ref{section:method} we present the method for producing the synthetic observations, as well as the observational datasets of the Milky Way used for this comparison. In Section~\ref{section:comparison_fiducial_obs} we present the results from our fiducial model, namely the $l-b$ and $l-v$ distributions of the CO and H{\sc{i}} emission, and the H$_{2}$ column densities, as well as the relation between the CO intensities, the H$_{2}$ column densities and the HISA. In Section~\ref{section:other_models} we explore the impact of different chemical and feedback recipes on these same results. Finally, in Section~\ref{section:conclusions} we summarise our findings and present our conclusions.


\section{Method}
\label{section:method}

The generation of a synthetic Galactic plane survey is a two stage process.  First, a smoothed particle hydrodynamics (SPH) model of a whole spiral galaxy is run to generate density, temperature, velocity and molecular abundance distributions (as described in Section~\ref{section:galaxy_model}). These results are then used as input to the {\sc{torus}} radiative transfer code \citep{Harries00} to generate synthetic observations in Galactic co-ordinates (as described in Section~\ref{section:RT_calculations}). The procedure for generating H{\sc{i}} observations is the same as that used by \cite{douglas_2010} and \cite{Acreman12}, but in this paper we extend the previous work by generating synthetic observations of the CO $(1-0)$ transition. The different sets of observations of our Galaxy that we will compare our models to are described in Section~\ref{section:observations}.

\subsection{The galaxy models}
\label{section:galaxy_model}

We have carried out a number of different models of galaxies which we use to make synthetic maps. All the simulations model the gaseous component of the galaxy, and adopt a gravitational potential to represent the dark matter halo, disc and a four armed spiral. The exact spiral structure of the Milky Way is unknown \citep[e.g.][]{vallee2014} although in the outer parts of the Galaxy, at least, there are likely more than 2 arms \citep[][]{englmaier2011,Pettitt2014}. Whether the spiral arms are truly long-lived (as assumed here) or transient, our four armed model does produce all the nearby features of the second quadrant (Perseus, Local and Outer Arm). The simulations all include heating and cooling of the ISM, as described in \citet{Glover07} and \citet{DGCK08}. Molecular hydrogen formation is included as described in \citet{Dobbs06} and \citet{DGCK08}. CO formation is included according to the prescription of \citet{Nelson1997}, and is also described in depth in \citet[][]{Pettitt2014}. In most of the simulations the gas is subject to self gravity, and we input stellar feedback where stars are assumed to form. The stellar feedback nominally represents supernovae, and is input in the simulations as described in \cite{Dobbs11b}. Specifically, the stellar feedback is inserted instantaneously as a combination of kinetic and thermal energy, according to a Sedov solution. The number of massive stars ($>8$M$_{\odot}$) formed, and therefore the energy inserted for each feedback event, is calculated from the mass of molecular gas in neighbouring particles (bound gas above a density of 500 cm$^{-3}$) multiplied by an efficiency (which is 5 per cent), and assuming a Salpeter IMF. All our simulations use 4 or 8 million particles which is nominally sufficient to capture CO emission \citep[][]{Pettitt2014}. However in all the simulations with stellar feedback, the feedback is inserted at a density of $\sim$500 cm$^{-3}$ and there is a minimum temperature of 50 K, so we bear in mind that these constraints could potentially limit CO formation. 

\begin{table}
\caption{Description of main SPH galaxy models used in this paper.}
\label{sph_models}
\renewcommand{\footnoterule}{}  
\begin{tabular}{l c c c c }
\hline \hline
Simulation & $\Sigma_{gas}$ & Self gravity  & $l_{ph}$$^*$ & No.  of \\
		& (M$_{\odot}$pc$^{-2}$)  & \& feedback &  & part. \\
\hline
Fiducial & 8 & Y & 35 & 8\,M \\
\hline
\multirow{2}{2.0cm}{Strong Self Shielding} & \multirow{2}{*}{8} & \multirow{2}{*}{Y} & \multirow{2}{*}{100} & \multirow{2}{*}{4\,M} \\
&&&&\\
\hline
\multirow{2}{2.0cm}{High Surface Density} &  \multirow{2}{*}{16} &  \multirow{2}{*}{Y} &  \multirow{2}{*}{35} &  \multirow{2}{*}{4\,M} \\
&&&&\\
\hline
No feedback & 8 & N & 35 & 8\,M\\
\hline
\end{tabular}
$^*l_{ph}$ is a measure of the column density used for self-shielding (see text).
\end{table}

We made synthetic emission maps for a total of eight different simulations, but focus on four simulations for this paper (summarised in Table~\ref{sph_models}). Our fiducial simulation is the same as that presented in \citet[][]{Dobbs2013} except this simulation uses a four-armed, rather than two-armed spiral potential \citep[a four-arm potential providing a better match to the ISM observations of the Milky Way, see e.g.][]{Pettitt2014}. It has a surface density of 8\,M$_{\odot}$pc$^{-2}$, uses 8 million SPH particles and the mass of each particle is 312.5\,M$_{\odot}$. We also present results from a simulation which is the same as that used for \citet[][]{douglas_2010}, except CO is now included. This simulation did not include stellar feedback or self gravity, hence we refer to it as the no feedback simulation (with the same particle mass as the fiducial model). In this model with no feedback there is no imposed limit for the temperatures and densities. A third simulation uses a higher surface density, of 16\,M$_{\odot}$\,pc$^{-2}$, but again includes self gravity and stellar feedback. With 4 million SPH particles, the mass of each particle is 1250\,M$_{\odot}$ in this case.

Finally, we performed another simulation which relates to the numerical implementation of H$_{2}$ and CO formation, as there are uncertainties about different parameters (e.g. formation efficiencies on grains) and in particular an approximation for calculating the photodissociation rate. To calculate photodissociation, we need to determine the degree of self shielding, which depends on the column density of molecular gas. We estimate the column density by multiplying the local density, $\rho$, by a length scale, $l_{ph}$. Our fiducial value of $l_{ph}$ is 35 pc \citep[see][]{DGCK08}. Here we test the sensitivity of this approximation by also comparing with a simulation with $l_{ph}$=100 pc, our strong self shielding model. With 4 million SPH particles, the mass of each particle in this model is 625\,M$_{\odot}$.

The remaining four models that we will not present in detail, were performed so as to check whether our implementation of feedback changed the amount of CO, and a summary of these can be found in Sect.~\ref{sec:other_sims}.

\subsection{Radiative transfer calculations}
\label{section:RT_calculations}

\begin{figure*}
	\hspace{-0.2cm}
  	\includegraphics[width=0.48\textwidth]{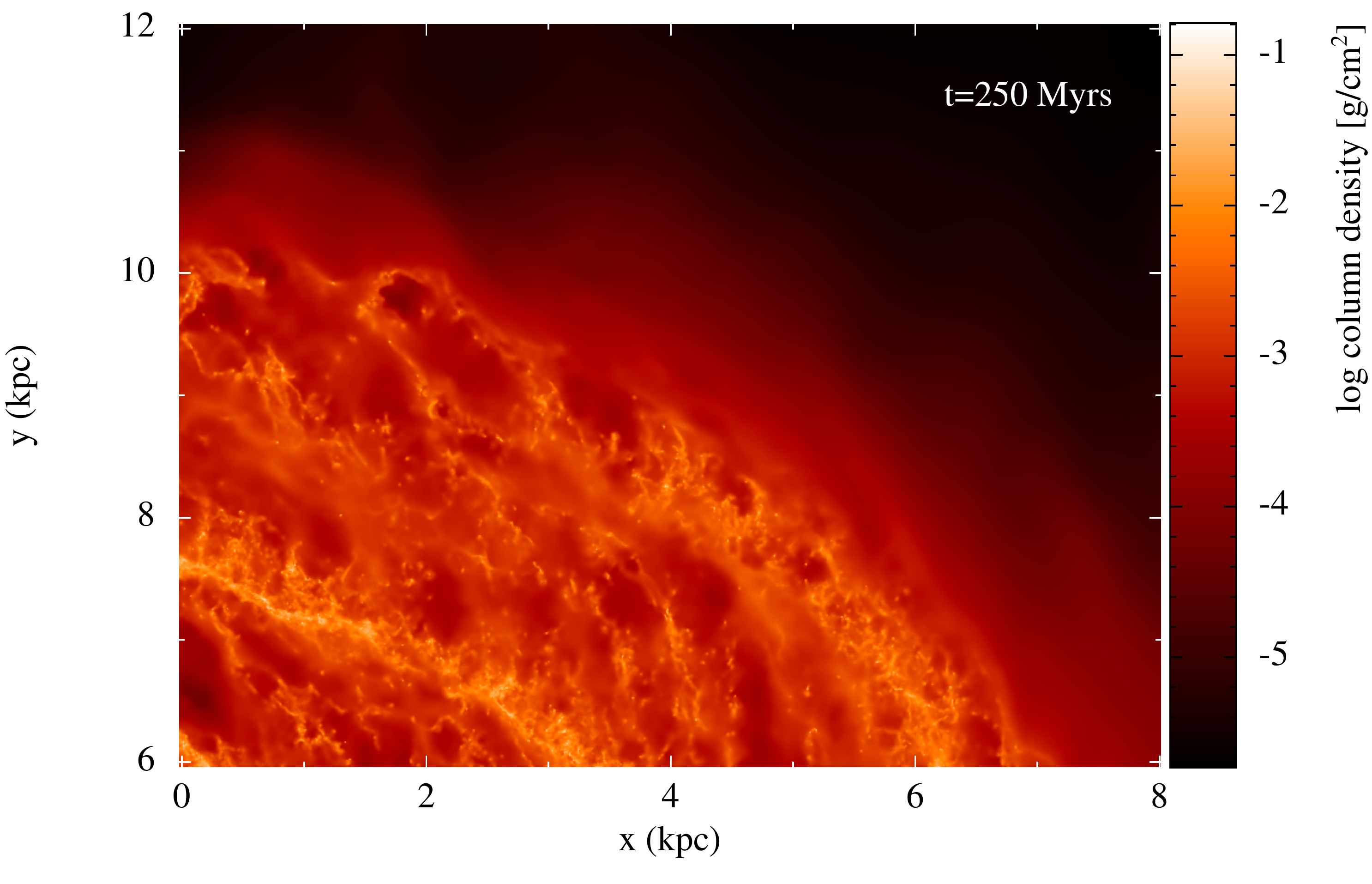} 
	\hfill		  		
	\includegraphics[width=0.48\textwidth]{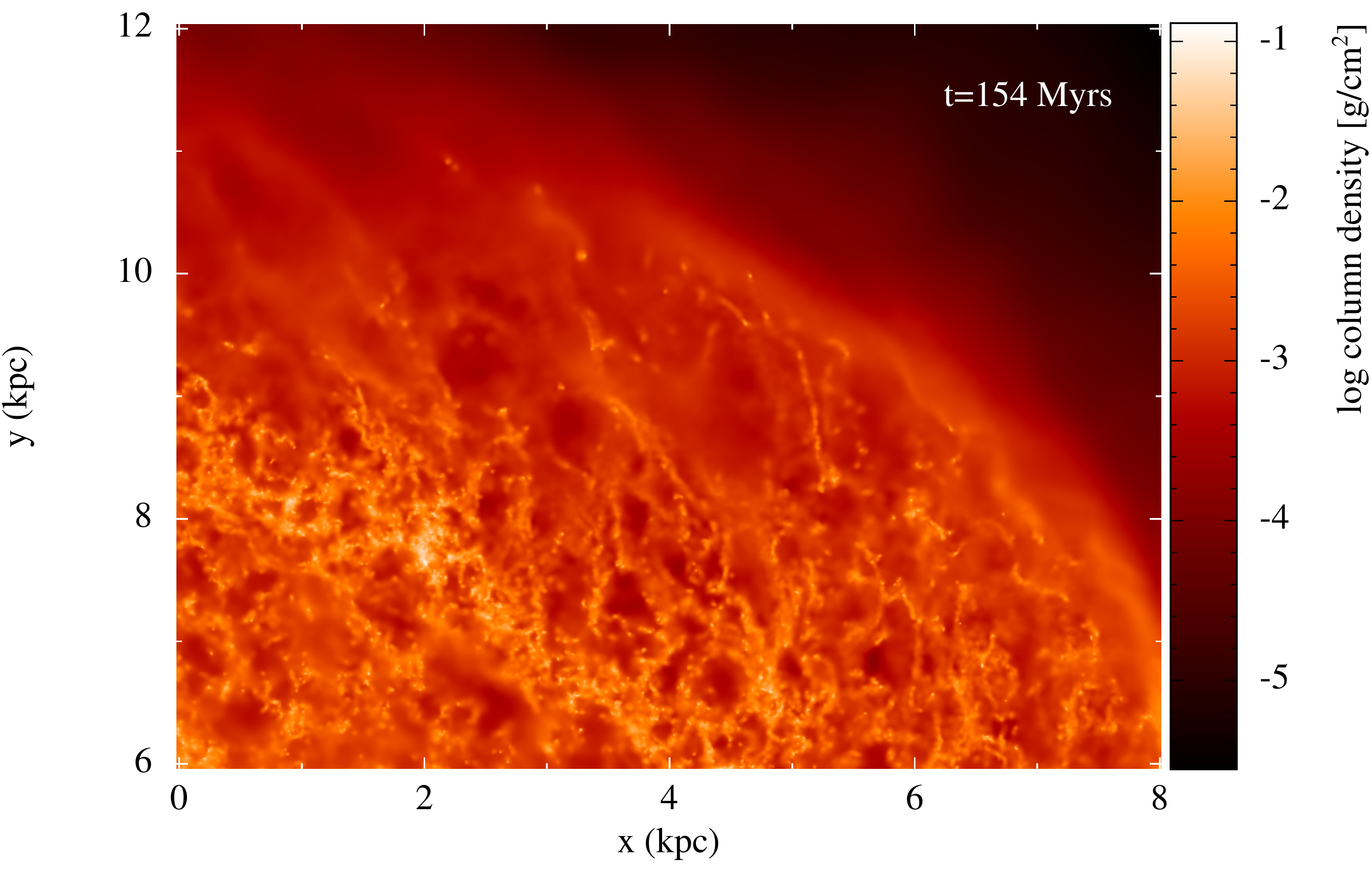} 
  	\caption{Top-down view of the column density structure of the second galactic quadrant from the fiducial model (on the left) and from the high surface density model (on the right). The observer is positioned on the bottom-left corner (at $x=0$\,kpc and $y=6$\,kpc), and sits on a ``local'' arm. The nearest arm, also the stronger, is the equivalent of the Perseus arm, sits between 2-4\,kpc distance, while the fainter Outer arm is situated at 4-7\,kpc distance from the observer.}
	\label{fig:top-down}
\end{figure*}

We have produced synthetic observations by placing the observer inside the simulated galaxies, so as to observe the emission equivalent to the second quadrant of our Galaxy (which we will abbreviate to 2Q hereafter). This implied positioning the observer so that the second quadrant had two arms which would lie at a distance corresponding to the stronger Perseus arm (between $\sim$2-3\,kpc), and the weaker Outer arm (between $\sim$4-6\,kpc). A top-down view of the 2Q from the fiducial model is shown in Fig.~\ref{fig:top-down}.

The synthetic H{\sc{i}} observations are then generated using the method described in \cite{douglas_2010} which produces spectral cubes in Galactic latitude-longitude-velocity co-ordinates. In addition to calculating H{\sc{i}} brightness temperature we also explicitly calculate the absorption component to determine H{\sc{i}} self-absorption (HISA). The spectral cubes have velocity channels of 0.5\,km\,s$^{-1}$ (over a velocity range of $-100$ to $+10$~km/s) and a pixel size of 1$'$. 

The synthetic spectral cubes of CO ($1-0$) emission are calculated using the molecular physics module of {\sc{torus}} as described in \cite{rundle_10}, which maps the SPH simulation into an AMR grid. Two CO ($1-0$) datacubes are generated one with and one without making the assumption of local thermodynamic equilibrium (LTE) and the large velocity gradient (LVG) approximation \citep[e.g.][]{santander_2012}. The latter requires calculating non-LTE level populations of the CO molecule in each cell of the AMR grid.  Once the level populations have been determined the emissivity and opacity of each cell on the AMR grid can be calculated and a spectral cube of CO emission generated using the same ray tracing method as used for generating H{\sc{i}} cubes. 

Even though the morphology of CO emission is the same, irrespective of whether LTE is assumed, the brightness temperature is affected by the LTE assumption. For determining the location of CO emission it is acceptable to assume LTE, but for retrieving the correct brightness of the emission a non-LTE calculation is required, as the line intensities are generally overestimated when assuming LTE. However, due to the computational effort this requires, the non-LTE spectral cubes of CO ($1-0$) emission were only generated for the sub-set of four galaxy models shown in Table~\ref{sph_models}, out of the eight mentioned in Sect.~\ref{section:galaxy_model}. 

The last step to mimic real observations requires the introduction of noise into the datasets and convolution with a 2D-gaussian representing a telescope beam. In this case, they were convolved with a 2D-gaussian of 4-pixels FWHM (equivalent to a 4$'$ beam), similar to the lowest resolution observations we are using for comparison (see Sect.~\ref{section:observations}). The resulting datasets of CO have a noise r.m.s. of $\sim$0.08\,K (in 0.5\,km\,s$^{-1}$ channels), comparable to the noise in the FCRAO dataset we compare to. The final synthetic H{\sc{i}} and HISA datacubes have a noise r.m.s. of $\sim$0.3\,K, equivalent to the noise in the CGPS H{\sc{i}} data after smoothing to a 4$'$ resolution.

Finally, we also generated column density maps of H$_{2}$, by mapping the SPH particles onto the AMR grid and tracing a path through the grid. This yields no velocity information but can be used to determine the distribution of H$_{2}$ as seen on the plane of the sky, comparable to the column density maps of our Galaxy as constructed from the observed dust continuum emission. The synthetic H$_{2}$ column density maps do not require noise addition since the observed map is a reconstructed map from an SED fitting. They have, however, been equally convolved with the same 2D-gaussian to reproduce the resolution of the observed $N$(H$_{2}$) map.

\begin{figure*}
\flushleft
  	\includegraphics[height=5.7cm]{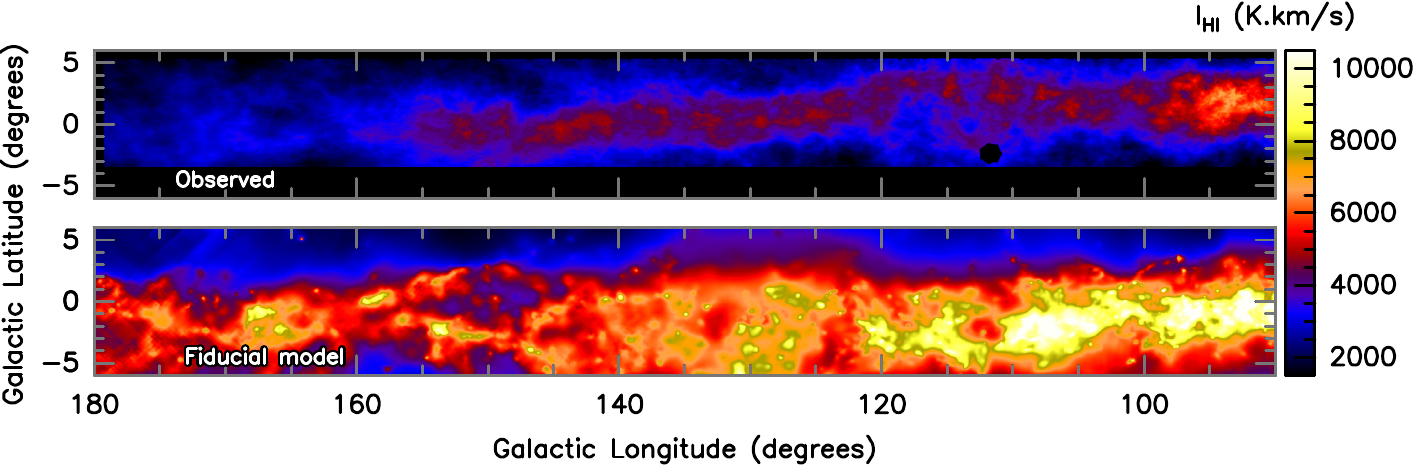} 
  	\includegraphics[height=5.77cm]{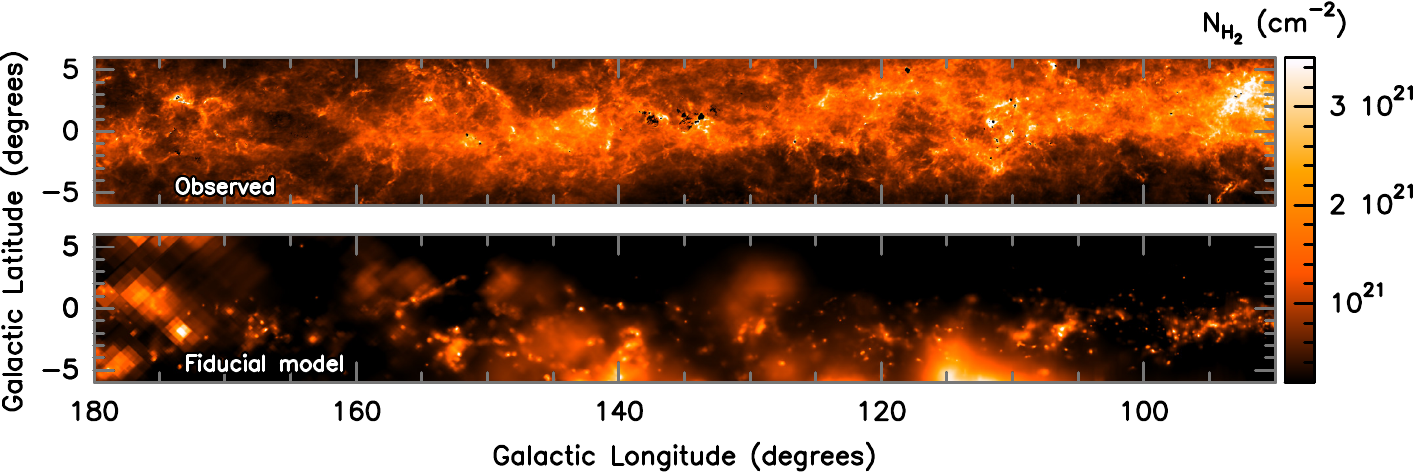} 
  	\includegraphics[height=5.7cm]{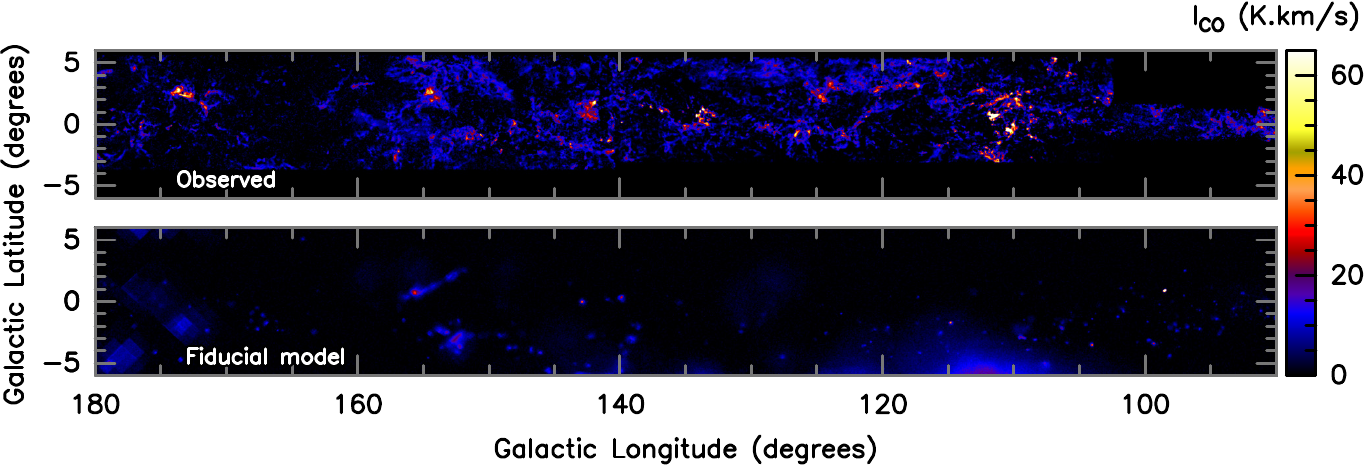} 
  	\caption{Comparison of the spatial distribution and intensities of the observed Galactic 2Q, and the galactic fiducial model. The H{\sc{i}} integrated emission is shown on the top, the H$_{2}$ column densities in the middle, and $^{12}$CO integrated intensities in the lower panels.}
	\label{fig:fiducial_spatial}
\end{figure*}

\subsection{Observations of the Milky Way}
\label{section:observations}

For comparing with the models, we used observations of the 2Q of the Milky Way in $^{12}$CO, H{\sc{i}} (and HISA) and dust continuum emission (for reconstructing the H$_{2}$ column densities).

The H$_{2}$ column density map was created by reconstructing a single grey-body SED on a pixel-by-pixel basis using dust continuum emission observed with IRAS at 100\,$\mu$m, with a beam size of 4.7$'$ FWHM \citep[][]{Wheelock1994}, and Planck at 350\,$\mu$m and 850\,$\mu$m, with beam sizes of 4.3$'$ and 4.8$'$ FWHM \citep[][]{planck2014}. These data were retrieved from SkyView\footnote{http://skyview.gsfc.nasa.gov}, covering $90^{\circ} \le l \le 180^{\circ}$ and $-6^{\circ} \le b \le 6^{\circ}$, and all resampled onto a pixel size of $1'$. For the SED fittings, we assumed an opacity law as in \citet[][]{Hildebrand1983} with $\beta = 2$, a dust emissivity of 1.0\,cm\,g$^{-1}$ at 1.3mm \citep[][]{Ossenkopf1994} and a dust-to-gas ratio of 100. Because we are only interested in retrieving the column densities of the cold molecular gas, we made use of IRAS 60\,$\mu$m data \citep[with a beam size of 3.6$'$ FWHM,][]{Wheelock1994}, which is a good tracer of hot ionised regions or point-like stellar objects and often includes emission from stochastically-heated very small grains, to discriminate between warm and cold material. In practice, we use only the column densities at positions that have a well constrained SED fitting (with low uncertainties), that have temperatures below 22\,K, and whose 60\,$\mu$m emission is sufficiently weak (below 300\,MJy\,sr$^{-1}$), so that the SED is properly fit by a single grey-body function. The excluded regions can be seen as small black patches in the observed $N$(H$_{2}$) map (top panel of middle row of Fig.~\ref{fig:fiducial_spatial}). 

We note that the gas column densities derived from the thermal dust continuum emission are, in fact, the total column densities of hydrogen, $N$(H$_{\rm total}$), and include a contribution from both molecular and atomic gas. \citet{McKee2010} estimate that the typical turn-over between atomic- and molecular-dominated gas occurs at $\sim$10-20\,M$_{\odot}$/pc$^{2}$, which corresponds to $\sim5-10\times10^{20}$\,cm$^{-2}$. Since we only use the column densities derived at positions where there is already significant molecular line emission (from CO), and this corresponds to relatively high total column densities ($>5\times10^{20}$\,cm$^{-2}$), we will be assuming that the amount of molecular gas dominates over the atomic gas, and hence the dust-derived column densities are assumed to be a good proxy of the molecular column densities. We consider that this assumption could result in over-estimating $N$(H$_{2}$) by up to a factor of two \citep[e.g.][]{gir1994}.

\begin{figure*}
\flushleft
  	\includegraphics[height=5.7cm]{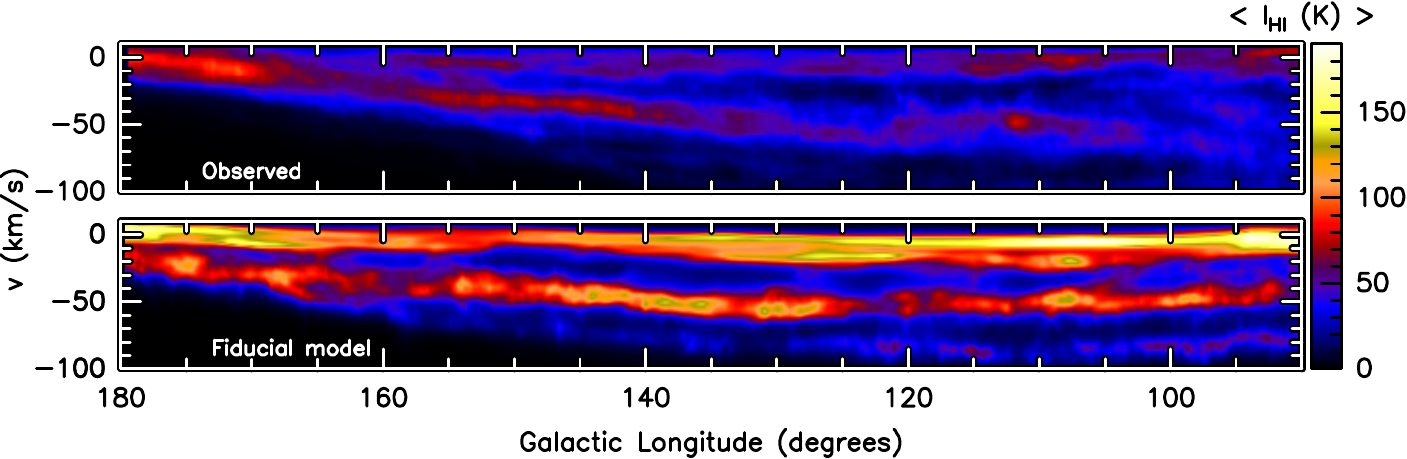}  
  	\includegraphics[height=5.7cm]{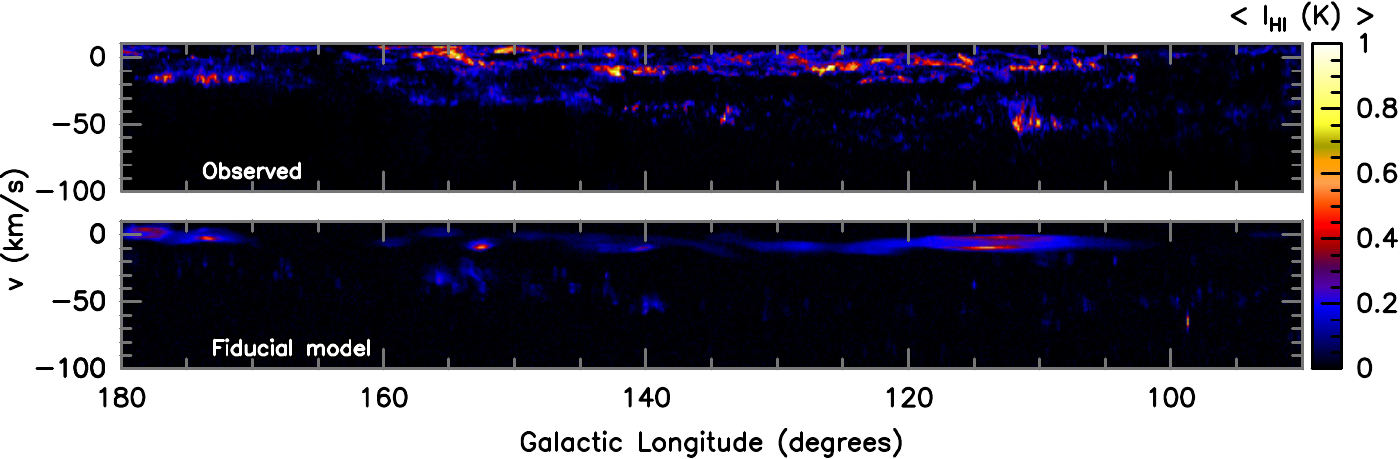} 
  	\caption{Comparison of the velocity distribution and average intensities of the observed Galactic 2Q, and the galactic fiducial model, for the H{\sc{i}} shown on the top, and the $^{12}$CO emission in the lower panels.}
	\label{fig:fiducial_velocity}
\end{figure*}

The $^{12}$CO (1-0) observations are a combination of the Exeter FCRAO CO Galactic Plane Survey \citep[][Brunt et al. in prep, covering from $90^{\circ} \le l \le 104^{\circ}$ with $b=\pm 1^{\circ}$, and from $135^{\circ} \le l \le180^{\circ}$ with $-3^{\circ} \le b \le 5^{\circ}$]{Mottram2010}, and the FCRAO CO Outer Galaxy Survey \citep[][covering between $104^{\circ} \le l \le 135^{\circ}$, with $-3^{\circ} \le b \le 5^{\circ}$]{Heyer98}. The original spatial resolution of these surveys is $45''$, and the spectral resolution is $0.15$\,km\,s$^{-1}$. Full details on the specific observations can be found in the respective survey papers. For the purpose of this study, we resampled the data to 1\,km\,s$^{-1}$ velocity channels, and to $1'$ pixels. Finally, we convolved the datacube to an equivalent beam of 4$'$ (to have the same spatial resolution as the H$_{2}$ column density map). The noise of resulting datacube is $\sim 0.07$\,K. The $^{12}$CO integrated intensity map across the entire velocity range is shown in Fig.~\ref{fig:fiducial_spatial} (top panel of bottom row).

Finally, the H{\sc{i}} datacube, covering $90^{\circ} \le l \le 180^{\circ}$ and $-3.6^{\circ} \le b \le 5.6^{\circ}$, is part of the Canadian Galactic Plane Survey \citep[][]{Taylor03}, and has an original angular resolution of 1$'$. The integrated intensities of the H{\sc{i}} emission across the entire velocity range can be seen in Fig.~\ref{fig:fiducial_spatial} (top panel of top row). The HISA datacube was produced from this H{\sc{i}} dataset using the methods described by \citet{Gibson05}. We have resampled both the H{\sc{i}} and HISA datacubes so as to have the same spectral and spatial sampling as the CO data, that is, 1\,km\,s$^{-1}$ velocity channels and $1'$ pixels, and convolved to an angular resolution of 4$'$. The noise in the resulting datacubes is $\sim$\,0.3\,K.


\section{Comparing the Fiducial Model with Observations}
\label{section:comparison_fiducial_obs}

\subsection{Spatial distribution}
\label{section:fiducial_lb_distribution}

Figure~\ref{fig:fiducial_spatial} (top row) shows the spatial distribution of H{\sc{i}} emission from the observations (above) and the fiducial model (below). From this comparison, we can see that the model is able to reproduce the latitude extent of the Galactic H{\sc{i}} (see also Sect.~\ref{section:lb_distributions} and Fig.\ref{fig:lat_dist} that discuss the average latitude distributions for all the models). 
Note, however, that the observed distribution of material is offset to higher latitudes, whilst the fiducial model is slightly shifted to lower latitudes. The observational shift is perhaps due to the Galactic warp, which is not included in the simulations. On the other hand, the only process in the models capable of pushing material off the midplane of the disc is the feedback. Hence, the shift of the fiducial model to lower latitudes, is purely by chance, and a consequence of the chosen observer position and the specific time frame. However, we consider that such shifts are not an issue for the work we present here, as we simply compare statistical properties of the ISM.
From Fig.~\ref{fig:fiducial_spatial} (top row) we can also see that for both observations and simulations there are numerous dense clouds of atomic hydrogen, in addition to other small scale ISM structures, and it is in these dense clouds where we expect atomic hydrogen to be converted to the molecular phase. There is more structure apparent in our fiducial model, compared to that of \cite{Acreman12}, due to the increase in resolution of the SPH model (8~million particles rather than 1~million particles), and the changed observer position so as to catch not only the Perseus Arm, but also the fainter Outer Arm.

The middle row of Figure~\ref{fig:fiducial_spatial} shows the distribution of molecular hydrogen column densities from the observations (above) and the fiducial model (below). The observations show low column-density widespread filamentary structures, that we do not recover in the simulations, due to insufficient resolution. Because low density material is represented by few (larger) particles, we cannot resolve any low-density structures smaller than the smoothing length of such particles. Similarly to the H$_{2}$ column densities, the CO emission (Fig.~\ref{fig:fiducial_spatial}, lower row) appears to be significantly less structured and more compact in the models than in observations. 
Furthermore, specially at the higher densities, we note that some of the clouds' internal sub-structures may be missing simply due to the fact that our models do not have the resolution to trace star formation or any other processes occurring at sub-parsec scales, nor do they include magnetic fields. In particular, the radiative feedback from young massive stars could potentially be important, as it can change the morphology of the native clouds up to tens-of-parsec scales, an effect that could therefore be visible at the resolution of the fiducial model. Nevertheless, we believe these limitations would be mostly ``cosmetic'' rather than capable of changing the global properties of the ISM.  
The range of recovered H$_{2}$ column densities of the fiducial model is similar to the observed one, but the spatial distribution (in latitude-longitude) is relatively sparse for the simulations with respect to the observed. Furthermore, the fiducial model tends to overestimate the H{\sc{i}} intensities, and underestimate the CO emission, which could indicate that the conversion of atomic gas onto molecular gas in this particular simulation is not sufficiently efficient. However the distribution of H$_{2}$ versus density is similar to that shown in \citet[][]{DGCK08}, the transition from H onto H$_{2}$ occurring at a few cm$^{-3}$. The difference here is that there is less gas at higher densities.

We stress that the fact that we do not probe densities much higher than 500 cm$^{-3}$, nor temperatures below 50K, likely results in an under-production of molecular gas (H$_{2}$), and consequently, also less CO is able to form. We believe this is in fact the main reason why the CO and H$_{2}$ are so sparse and compact, as there is simply not enough molecular material to be able to trace neither the full extend nor the finer details of clouds. Furthermore, the minimum temperature of 50K for the models is on the warm end of what we actually observe with the CO FCRAO data (only the bright compact regions get above T$_{ex}\sim30$\,K). This likely leads to more molecules populating higher excited states in the models, resulting in lower CO\,$(1-0)$ intensities than what we would obtain for colder temperatures. This could be part of the reason why the CO\,$(1-0)$ emission is weaker in the models.

\subsection{Velocity structure}
\label{section:fiducial_lv_distribution}

Figure~\ref{fig:fiducial_velocity} shows the average H{\sc{i}} and CO intensities in longitude velocity space. The top panels show the observations of second quadrant of the Milky Way, and the second row shows the synthetic observations of the fiducial model. In the observations we can identify, particularly in H{\sc{i}}, three distinct velocity structures, which correspond to local material (that we will refer to as Local arm, at $\sim$0\,km\,s$^{-1}$ velocities), the Perseus arm (starting at $\sim -50$\,km\,s$^{-1}$), and the fainter Outer arm (reaching $\sim -100$\,km\,s$^{-1}$). The CO observations show emission in the same longitude-velocity space, except for the Outer Arm, whose emission is too faint to be detected. The fiducial galaxy model is able to trace the same features, and the correlation is quite remarkable in terms of both distribution and relative intensities of H{\sc{i}}, even though the fiducial model overestimates the absolute intensities of H{\sc{i}} emission. For the CO, the model reproduces the equivalent of the Perseus Arm, though the emission is more compact than observed. The model also has some emission from local material which is, however, substantially unresolved (hence the smooth large-scale appearance). As in observations, the outer arm is undetectable with CO in the model.

\subsection{CO and $\rm{H}_2$ column densities}
\label{section:fiducial_column_densities}

\begin{figure}
\centering
 \includegraphics[width=0.45\textwidth]{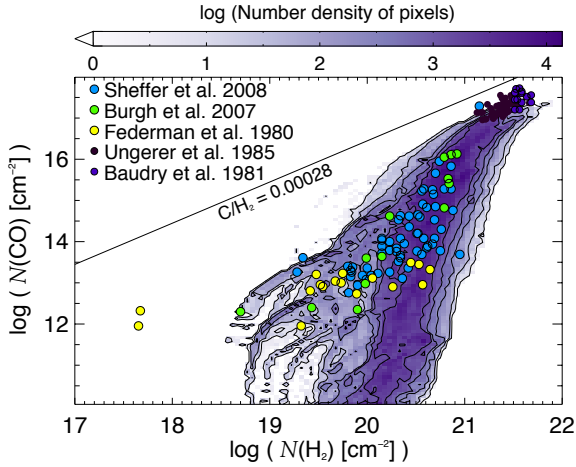}
\caption{Distribution of the CO column density against the $\rm{H}_2$ column density from Galactic observations in the literature (coloured circles) and from the fiducial galaxy model (contours and colour scale). As in \citet[][]{sheffer_2008}, the line with C/H$_{2} = 2.8 \times 10^{-4}$ shows the observational limit for the formation of CO based on the supply of atomic carbon (i.e. the abundance of C/H) from \citet[][]{Cardelli1996}.}
\label{fig:fuse_fiducial}
\end{figure}

Observational determinations of Galactic molecular column densities, including $\rm{H}_2$ and CO, were made by, e.g., \citet[][]{sheffer_2008},  \citet[][]{Burgh07},  \citet[][]{Ungerer85}, \citet[][]{Baudry81}, and \citet[][]{Federman80}. We use these observational data to validate the line of sight column densities derived from our simulation. $\rm{H}_2$ and CO column densities from our synthetic observations of the fiducial model are plotted in Fig.~\ref{fig:fuse_fiducial} (contours and colour scale) with the observational values overlaid (coloured circles). From the figure we can see that the column densities derived from our fiducial model compare favourably with the observational column densities found by \cite{sheffer_2008} in diffuse clouds. For the higher column density clouds (above $10^{21}$\,cm$^{-2}$) the simulations yield marginally lower CO column densities, most likely as a consequence of the relatively low maximum volume densities and relatively high minimum temperatures in the simulations, that limit the production of CO at high densities.

\subsection{The relation between CO intensities and $\rm{H}_2$ column densities: the X$_{\rm CO}$ factor}
\label{sec:Xfactor_fiducial}

\begin{figure}
\centering
\includegraphics[height=6cm]{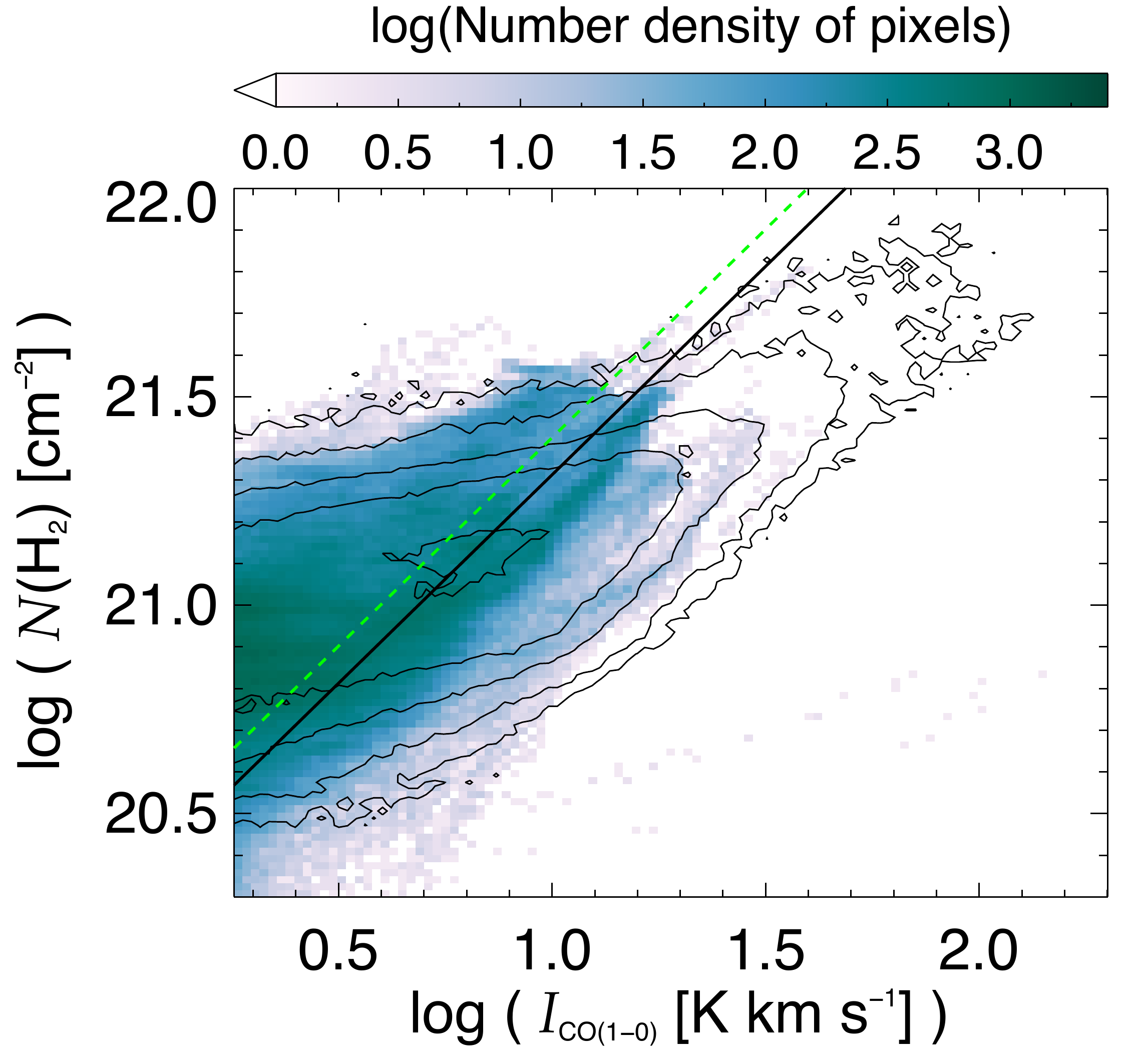}
\caption{Distribution of the CO intensities against the $\rm{H}_2$ column density from the observations (in black contours) and the fiducial model (colour scale). The green dashed line shows the median X$_{\rm CO}$ factor from the model, and the black solid line corresponds to the median observed X$_{\rm CO}$. This plot only shows emission above a $3\sigma$ noise level on the CO integrated intensities ($\sigma \sim 0.6$\,K\,km\,s$^{-1}$), and all pixels below this are not considered for estimating the X$_{\rm CO}$ factor.}
\label{fig:CO_nH2_Fiducial}
\end{figure}

\begin{figure}
\centering
\includegraphics[height=6cm]{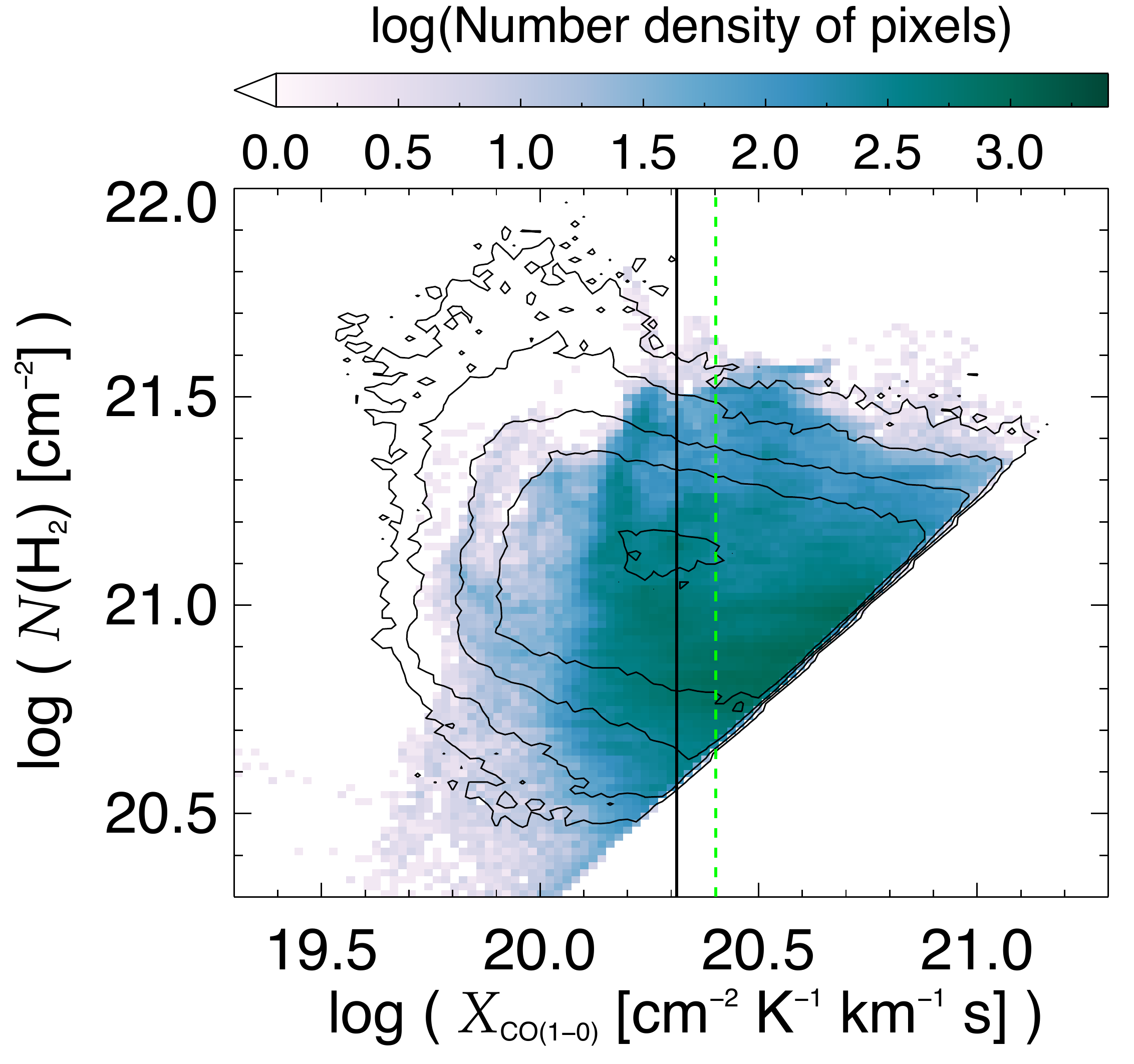}
\caption{Distribution of the X$_{\rm CO}$ factor against the $\rm{H}_2$ column density from the observations (in contours) and the fiducial model (colour scale). The black solid line shows the X$_{\rm CO}$ factor estimated from the observations, and the green dashed line shows the X$_{\rm CO}$ for the fiducial model. The diagonal cut-off on the right-hand side of this plot is due to the $3\sigma$ noise limitation of the CO integrated maps.}
\label{fig:XCO_nH2_Fiducial}
\end{figure}

Observationally, the $\rm{H}_2$ column density can be estimated using dust continuum emission or IR extinction. However, such observations are not always available, in which case $N(\rm{H}_2)$ can then be inferred from CO observations using a conversion factor termed the 'X-factor'. The $X_{\rm CO}$ factor is the conversion factor between CO brightness and the H$_{2}$ column densities. It is, therefore, defined as: $X_{\rm CO} = N({\rm H_{2}}) / I_{\rm CO}$, where $I_{\rm CO}$ is the CO integrated intensity. This factor has been estimated to be relatively constant in molecular clouds of our Galaxy \citep[with $X_{\rm CO} \sim 1.8 \pm 0.3 \times 10^{20}$\,cm$^{-2}$\,K$^{-1}$\,km$^{-1}$\,s, e.g.][]{Dame01}, and is therefore often used to directly convert from CO intensities to gas column densities. Despite being widely used, the $X_{\rm CO}$ factor is subject to caveats, as in practice the CO emission is not only a function of the gas column densities, but also a function of the gas temperature. Furthermore, the CO emission can become optically thick, in which case the $I_{\rm CO}$ saturates, while column densities may continue to increase. Therefore, $X_{\rm CO}$ is a valuable statistical quantity, but should be used with caution.
 
We have investigated the $X_{\rm CO}$ factor we recover in our galaxy models, as a test of whether we are able to form CO within the molecular gas in similar proportions to those observed, and also to investigate whether the $X_{\rm CO}$ factor changes with the column densities. 

Figure~\ref{fig:CO_nH2_Fiducial} shows the distribution of $N(\rm H_{2})$ column densities against CO integrated intensities. The colour scale and contours show the number density of pixels falling in each $I_{\rm CO}-N(\rm H_{2})$ bin. Most  X$_{\rm CO}$ factor calculations in the literature take the average $I_{\rm CO}$ and $N(\rm H_{2})$ within molecular clouds, therefore masking any internal variations of the X$_{\rm CO}$. As we have instead done a pixel-per-pixel comparison, we can see from this plot that there is in fact a significant spread resulting from the different conditions probed. From here, we have estimated the statistical median of the X$_{\rm CO}$ factors, using only the pixels where $I_{\rm CO}$ lay above $3\sigma$ of the noise ($\sigma \sim 0.6$\,K\,km\,s$^{-1}$). We estimated the respective scatter as the mean value of the absolute deviations at the first and third quartiles of the X$_{\rm CO}$ distribution. For the observations, we obtained a X$_{\rm CO} \sim 2.0\,(\pm\,0.9) \times 10^{20}$\,cm$^{-2}$\,K$^{-1}$\,km$^{-1}$\,s, consistent (within uncertainties) with the observed value quoted in the literature. For the fiducial model we obtained a somewhat higher value of X$_{\rm CO}\sim 2.5\,(\pm\,0.9)\times 10^{20}$\,cm$^{-2}$\,K$^{-1}$\,km$^{-1}$\,s. As the observationally determined value of \cite{Dame01} was averaged over latitude we would expect the variations we see within an individual cloud to be averaged out; this accounts for the smaller spread in the Dame best fit value compared to the values we retrieve here.

Figure~\ref{fig:XCO_nH2_Fiducial} shows the distribution of $N(\rm H_{2})$ column densities against the X$_{\rm CO}$ estimated for each pixel, so as to investigate possible variations of the X$_{\rm CO}$ with $N(\rm H_{2})$. We find that there is no strong global trend, although there is a possible hint of a marginal decrease of the X-factor towards increasing column densities, consistent with an increasing CO to H$_2$ ratio as expected if CO-bright clouds are surrounded by an envelope of H$_2$ gas which is still CO-dark (in this regime, the H$_{2}$ can self-shield from the ISM, but CO does not yet have a high enough shielding column). However, the scatter on our plot is too large to consider this a robust result. Nevertheless, similar X$_{\rm CO}$ factors and a similar trend of X$_{\rm CO}$ with $N(\rm H_{2})$ had been retrieved by \citet{Shetty2011a,Shetty2011b} on their lower density clouds, with average column densities similar to the range probed here.

\subsection{The relationship between HISA, CO and $\rm{H}_2$}
\label{sec:HISA_CO_H2_fiducial}

\begin{figure}
\centering
\includegraphics[height=6.5cm]{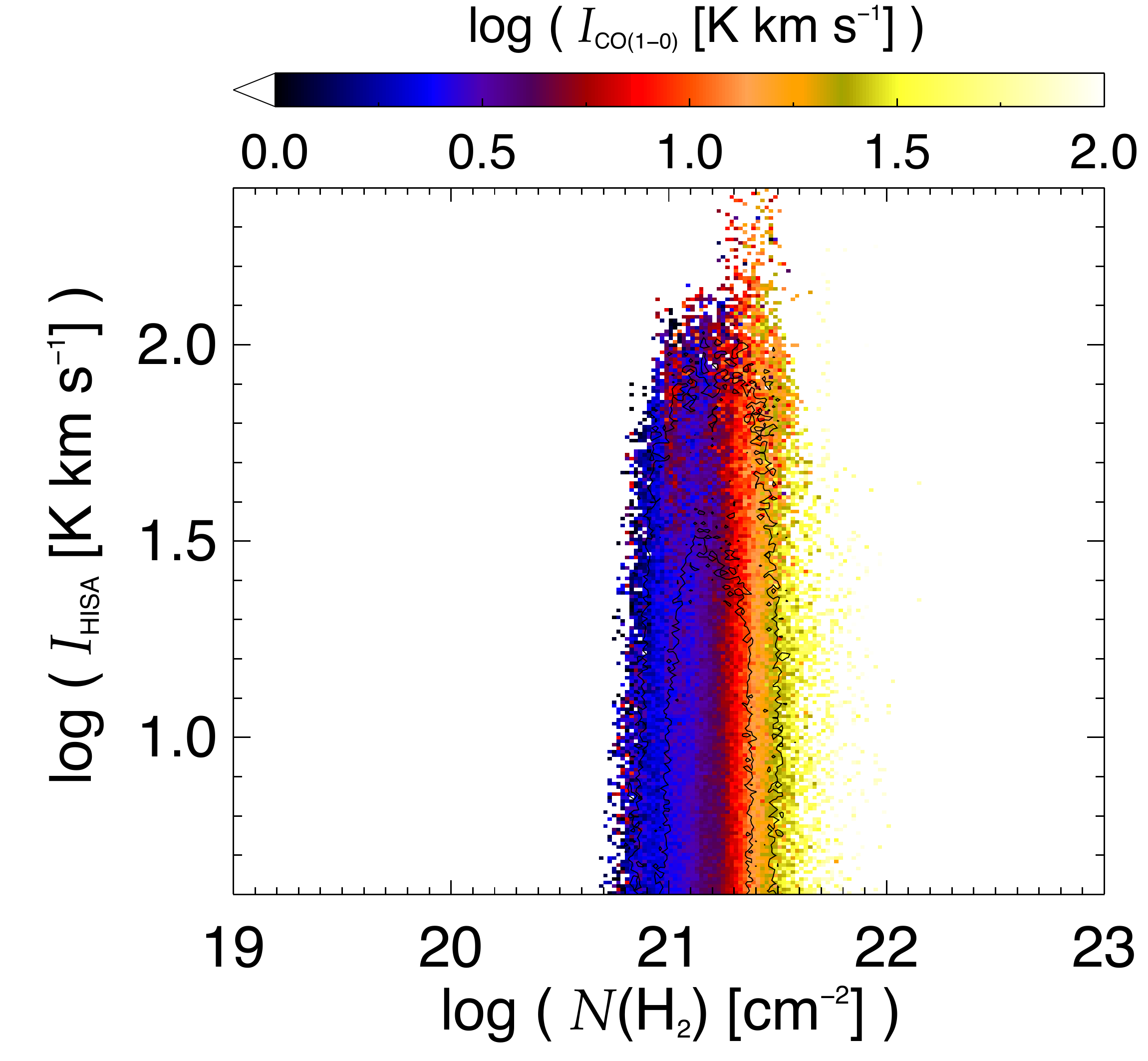}\\
\vspace{0.3cm}
\includegraphics[height=6.5cm]{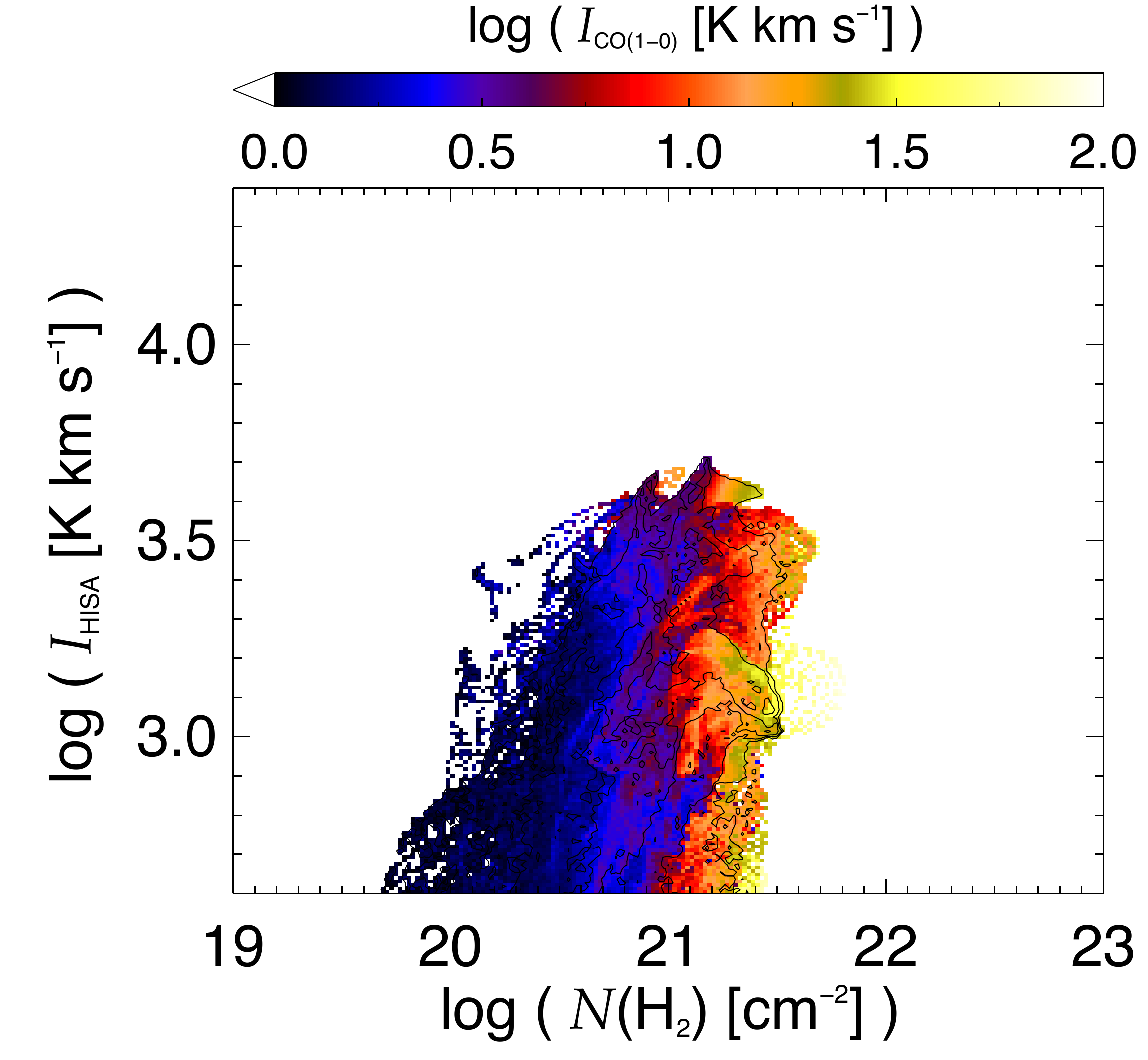}
\caption{Distribution of the absolute value of the HISA integrated intensities, against the corresponding $\rm{H}_2$ column densities, colour-coded with the CO integrated intensities, for the observations (top) and the fiducial model (bottom). The contours represent the density of pixels for each $I_{\rm HISA}$-$N(\rm{H}_2)$ bin.}
\label{fig:HISA_nH2_CO_fiducial}
\end{figure}

HISA is the self-absorption of the H{\sc{i}} emission from warm atomic hydrogen in the background, by colder atomic hydrogen along the line of sight. Therefore, HISA is expected to be correlated with the existence of cold molecular gas (H$_{2}$ and CO). However, this correlation has proven to be observationally hard to catch as the existence of HISA is not only dependent on the amount of cold atomic gas, but also the existence of warm background H{\sc{i}} emission to be absorbed \citep[c.f.][]{Gibson2000,Kavars2005}. Furthermore, observations have shown the existence of both HISA clouds with little molecular gas, as well as molecular clouds without HISA \citep[e.g.][]{Klaassen05,Gibson10}. 

We have investigated the statistical spatial relationship between HISA, CO and $N(\rm{H}_2)$, from the fiducial model, to understand if they behave in a similar manner in the simulations compared to observations of our Galaxy, and also, to look for any particular correlations. We did so by comparing the integrated intensities of HISA, H$_{2}$ column densities and CO intensities (see also Appendix~\ref{app:arms_obs}). Since our line of sight towards the second quadrant includes emission from the Local arm and the Perseus arm, it is important to do this comparison considering the integrated intensities of each of the arms separately. For instance, in the observations, while the CO emission is dominated by local emission, the HISA is more prominent in the Perseus arm. Therefore, we integrated the CO and HISA emission for the Local and Perseus arms, for both observations and model, for longitudes below 160$^{\circ}$ (after this, the velocities of both arms start to become degenerate). For longitudes below 160$^{\circ}$ we are able to separate the velocity ranges more easily, and as such, for the local emission, we integrated from 10\,km\,s$^{-1}$ to $-20$\,km\,s$^{-1}$, and for the Perseus arm we integrated from $-20$\,km\,s$^{-1}$ to $-70$\,km\,s$^{-1}$. For the $N(\rm{H}_2)$ we have no easy method to distinguish between the Local and Perseus arm, as we have no velocity information. Hence, we made use of the CO emission to evaluate if the CO integrated intensities from one arm dominate over the other, and attributed the entire $N(\rm{H}_2)$ to the dominating arm. If no dominant arm is distinguishable, we cannot attribute the column densities to any given structure, and as such do not use these positions. 

Figure~\ref{fig:HISA_nH2_CO_fiducial} shows a scatter plot of the absolute value of the HISA integrated intensities, against the corresponding $\rm{H}_2$ column densities, colour-coded with the CO integrated intensities. The observed HISA integrated intensities are significantly lower than those found in the fiducial model (by more than an order of magnitude). We believe this could be partially a consequence of the method used to obtain the observed HISA \citep[see][]{Gibson05} as the amplitude of the observed HISA only represents a lower limit to the actual intensity of the HI self absorption, while in simulations this is a direct output from the radiative transfer calculations, so it corresponds to the actual absolute HISA value. Another effect that could make the HISA intensities higher in the simulations could be the fact that we have too much atomic gas and too little H$_{2}$ and CO. If the amount of atomic gas with respect to molecular is too high, the simulations have not only more of the warmer atomic gas, but also more cold atomic gas which can absorb the background H{\sc{i}} emission, effectively increasing the HISA values.

From Fig.~\ref{fig:HISA_nH2_CO_fiducial}, we can see that the observations do not show any kind of correlation between $I_{\rm HISA}$ and the $N(\rm{H}_2)$, while for the fiducial model there is a tentative weak correlation between $I_{\rm HISA}$ and the $N(\rm{H}_2)$, though with an important dispersion. The CO intensities also seem to be rather uncorrelated with the amount of HISA, and they appear to depend solely on the amount of H$_{2}$. CO starts appearing at column densities above $\sim 10^{21}$\,cm$^{-2}$ for both observations and simulations. Whenever CO starts to be detected, it does not show any gradient of intensities along the y-axis, i.e. with $I_{\rm HISA}$, and instead, it varies nicely with the x-axis, i.e. with $N(\rm{H}_2)$. We note, however, that for these observations in particular, these results are affected by a number of effects: 1) there is significant HISA emission at $l > 160^{\circ}$ (which is not considered for this scatter plot); 2) there is also some strong HISA at $l \sim 93 ^{\circ}$ and $b \sim 3^{\circ}$, for which we have no CO coverage (and is therefore not considered for this plot). 3) there are some positions with strong $I_{\rm HISA}$ that coincide with positions that had been masked from our $N(\rm{H}_2)$ map. These issues are better seen in Appendix~\ref{app:arms_obs}, where we show the spatial distribution of the HISA and $N(\rm{H}_2)$ referent to the two arms (see Figs.~\ref{fig:local_perseus_ncol} and \ref{fig:local_perseus_hisa}). Therefore, the observational relation between HISA and $N(\rm{H}_2)$ (and CO) should be better explored in other galactic regions where we do not suffer from the coverage problems we have in the second quadrant (with the dataset we possess).

\subsection{Properties of molecular clouds}
\label{sec:cloud_properties_fiducial}

One other aspect that can be investigated is the properties of the molecular clouds formed and observed in CO within the fiducial galaxy model, and those of the Milky Way. To do so, we have extracted the clouds from the 3D datacubes (PPV) using the {\sc clumpfind} method of the Starlink {\sc findclumps} routine, using the same parameters for both observations and simulations. The parameters used in the detection algorithm were set so that the main body of the cloud was identified and not substructures within clouds. We start the detection whenever the CO emission is above $5\sigma$ of the r.m.s. noise level, and contours are spaced by $10\sigma$. We did not allow the detection of any clouds smaller than the beam size in the resampled map (of 4$'$). 

\begin{figure}
\centering
\includegraphics[width=0.45\textwidth]{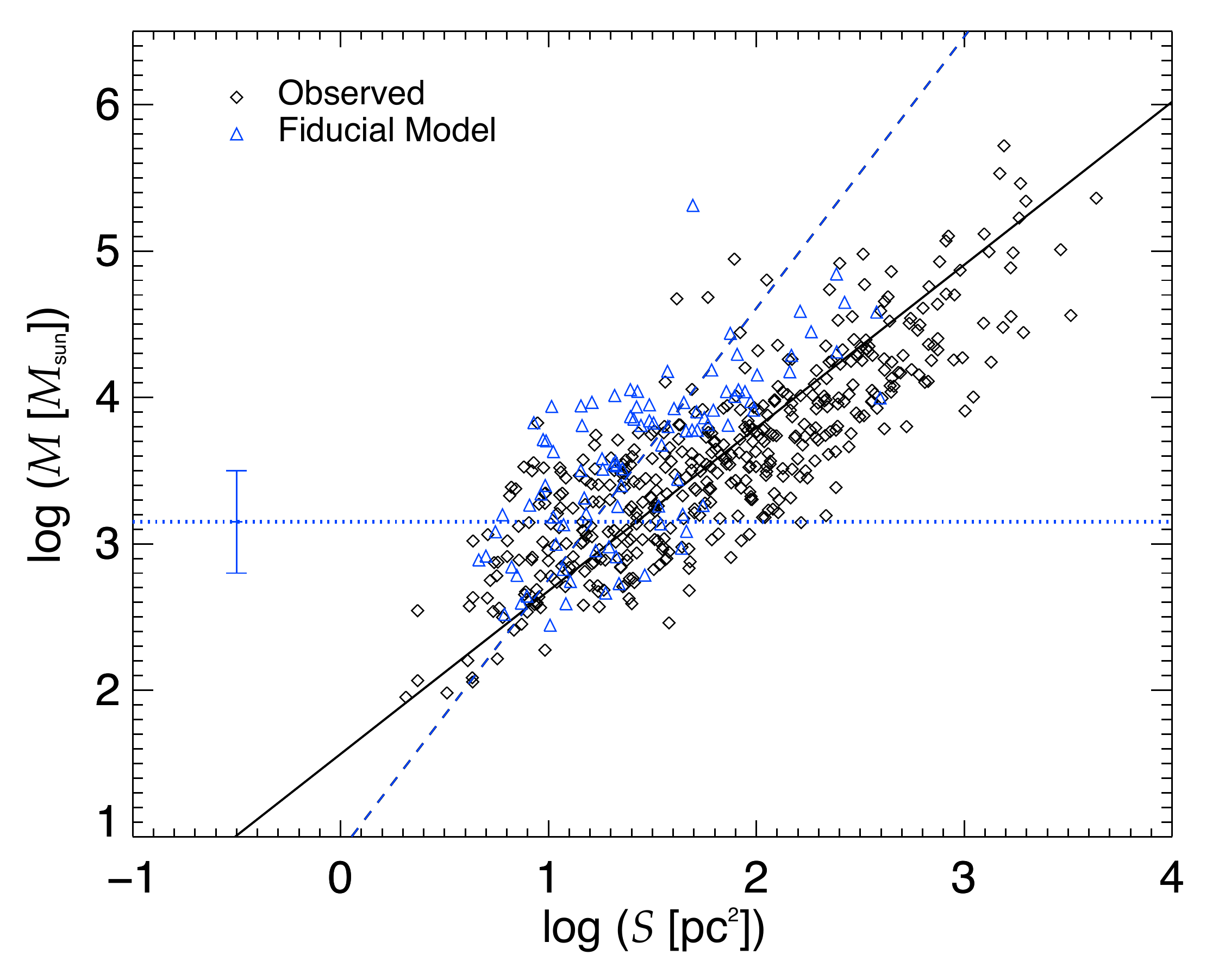}
\caption{Comparison of the cloud masses versus the cloud (projected) sizes as extracted from the CO emission, for the observations (in black diamonds) and for the fiducial galaxy model (in blue triangles). The respective linear fits to the log-log  space, corresponding to $M \propto S^{\alpha}$ are shown in black solid and blue dashed lines. Observations retrieve ${\alpha} = 1.1 \pm 0.1$ while the fiducial model has a steeper slope, with ${\alpha} = 1.8 \pm 0.6$. The blue dotted horizontal line indicates the average masses that we expect to be resolved, with an estimated level of the uncertainty shown with the error bar. }
\label{fig:larson_obs_fid}
\end{figure}

This extraction provides a catalogue of clouds for each map, from which we have estimated the masses and physical sizes. In order to do so, however, we need information on the distances. Even though for the fiducial model we could derive those directly from the simulation, we have chosen to adopt the same method as for the observations, for consistency. Therefore, we have calculated the kinematical distances for the clouds using the \citet[][]{2009ApJ...700..137R} galaxy rotation model for estimating the distances based on the velocities. We adopted a distance to the galactic centre of 8.4\,kpc for the observations, and of 6.0\,kpc for the fiducial model. We cross-checked the derived distances for the clouds of the galactic models with the actual distribution of clouds from the simulations, to assess the uncertainties of this method. We estimate that errors on the distances can range from $\sim$15\% up to a factor 2, most often being overestimated (which will consequently overestimate cloud masses). The high surface density model was the model that suffered the most severe over-estimations of distances. This high uncertainty could be partially due to the fact that the galactic kinematical model used was built so as to reproduce the velocity structure of our Galaxy, and is not precisely tailored to our models. In addition, because the velocities become close to 0 at a longitude of 180$^{\circ}$, the method for calculating the distances close to 180$^{\circ}$ becomes degenerate, and the estimated distances are unreliable. Therefore, we have included only the clouds detected between 90$^{\circ}$ and 160$^{\circ}$ latitude for both observations and simulations. We also excluded clouds for which the uncertainty estimated from the \citet[][]{2009ApJ...700..137R} galaxy rotation model was larger than a factor 2.

The masses were then calculated by transforming the $I_{\rm CO}$ into H$_{2}$ column densities, by applying the X$_{\rm CO}$ factor derived in Sect.~\ref{sec:Xfactor_fiducial}, and assuming a molecular weight of 2.8. Figure~\ref{fig:larson_obs_fid} shows the Larson relation between size (i.e. the area of an ellipsoid with semi-major and semi-minor axes as from the {\sc{clumpfind}} output) and mass of the clouds, for the sample of clouds we have extracted from the observations, and from the fiducial model (see also Sect.~\ref{sec:molecular_by_CO} for a discussion about the cloud mass and size distributions for the observations, and the four models we study in the article). For the simulation, the resolution of the SPH model may limit the completeness of this plot. However the minimum resolvable mass is not readily determinable since different particles contribute different amounts of CO emission (although the mass determined from CO will always be less than the mass in particles). 
In fact, for the fiducial model, we have determined that the typical amount of molecular mass in the extracted clouds represents only $\sim10\%$ of the total cloud mass, and furthermore the molecular mass may not be entirely observable with CO. If we assume that a cloud is resolved when it is comprised of at least 50 SPH particles, that sets a lower limit of $\sim1.5\times10^{4}$\,M$_{\odot}$ for the total cloud mass, and therefore, $\sim1.5\times10^{3}$\,M$_{\odot}$ for the molecular cloud mass.
There is also an uncertainty arising from our estimate of X$_{\rm CO}$ (around 40$\%$), and the uncertainty from using the kinematical distances. Furthermore, the inclusion of noise in the dataset, will introduce a further uncertainty on the total flux which is recovered from any given cloud (as portions of the cloud will be below the noise level, and hence, not detected). Taking into account these various factors, we include in Fig.~\ref{fig:larson_obs_fid} a line indicating where on average masses we expect to be resolved lie, with an estimated level of the uncertainty.

We can see that the extracted clouds retrieve a linear trend on a log-log plot, which implies $M \propto S^{\alpha}$, where $S$ is the projected size (area) of the cloud. The observations retrieve the expected Larson law with $\alpha = 1.1\pm0.1$, while the fiducial model retrieves $\alpha = 1.8 \pm0.6$, which is still reasonably in agreement with the observations, within the uncertainties.


\section{Sensitivity of models to different chemistry/feedback recipes}
\label{section:other_models}

The results from the comparison of our fiducial model with the observations have shown that we are overproducing atomic gas with respect to molecular gas in the simulations. This could either be because we are not efficient enough at transforming atomic H into H$_{2}$, or because we are too efficient at destroying the molecular gas with the implemented feedback. In this section, we test how sensitive the models are to the changes of the recipes for the formation and destruction of molecular material, by analysing the effects of changes in the chemistry and feedback recipes from a set of parallel models (described in Sect.~\ref{section:galaxy_model}).  

\subsection{Spatial distribution}
\label{section:lb_distributions}

\begin{figure}
	\hspace{-0.2cm}
  	\includegraphics[width=0.48\textwidth]{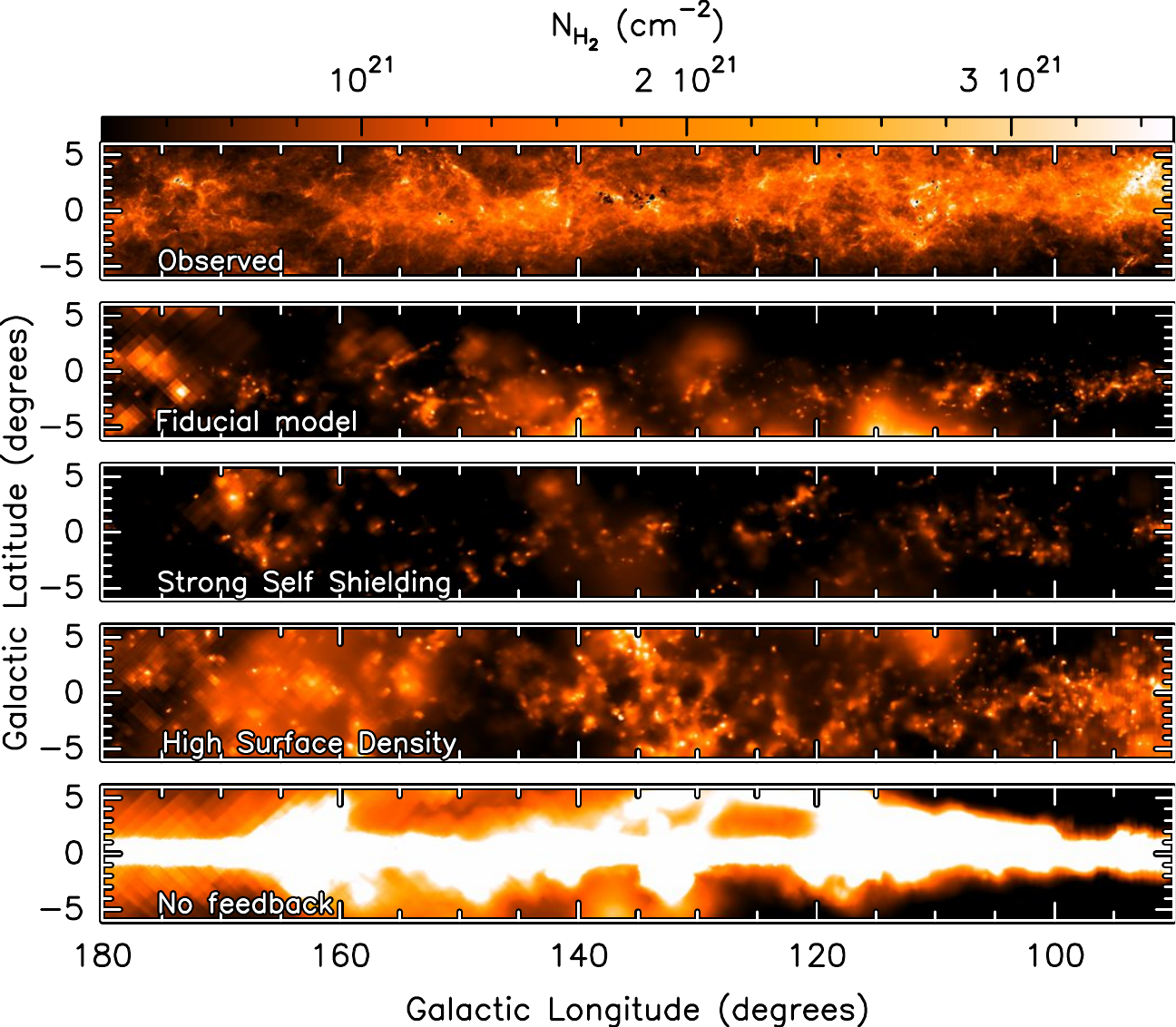} 
  	\caption{Comparison of the observed H$_{2}$ column density from the second quadrant of the Galaxy (top panel) and the galactic models (below).}
	\label{fig:spatial_coldens}
\end{figure}

\begin{figure*}
	\includegraphics[width=\textwidth]{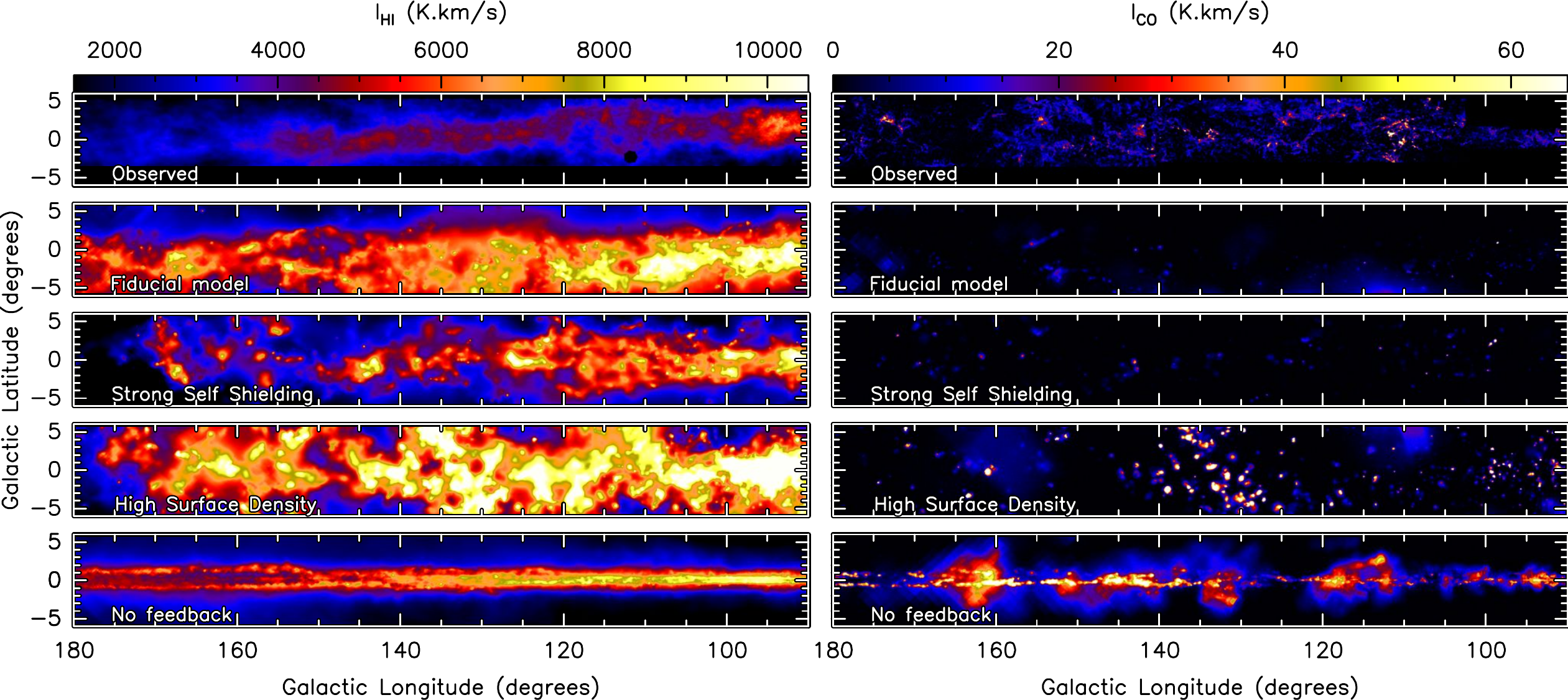} \\
  	\caption{Comparison of the observed l-b distribution of H{\sc{i}} ({\it left column}) and CO ({\it right column}) integrated intensity from the second quadrant of the Galaxy (top row) and the galactic models (on rows below). }
	\label{fig:spatial_HI}
	\label{fig:spatial_CO}
\end{figure*}

\begin{figure*}
	\hspace{-0.4cm}	
  	\includegraphics[width=0.36\textwidth]{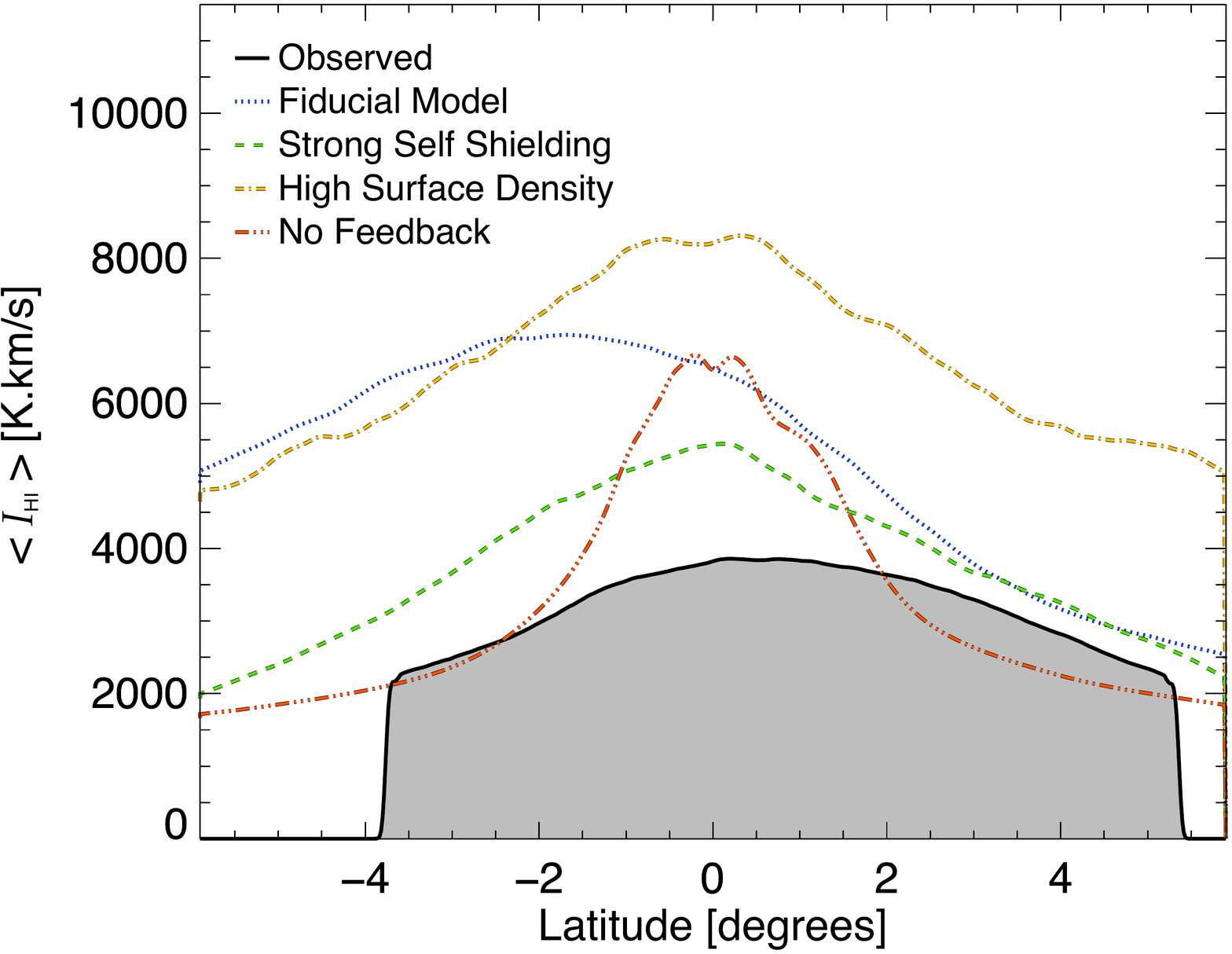}
	\hspace{-0.75cm}
	\includegraphics[width=0.36\textwidth]{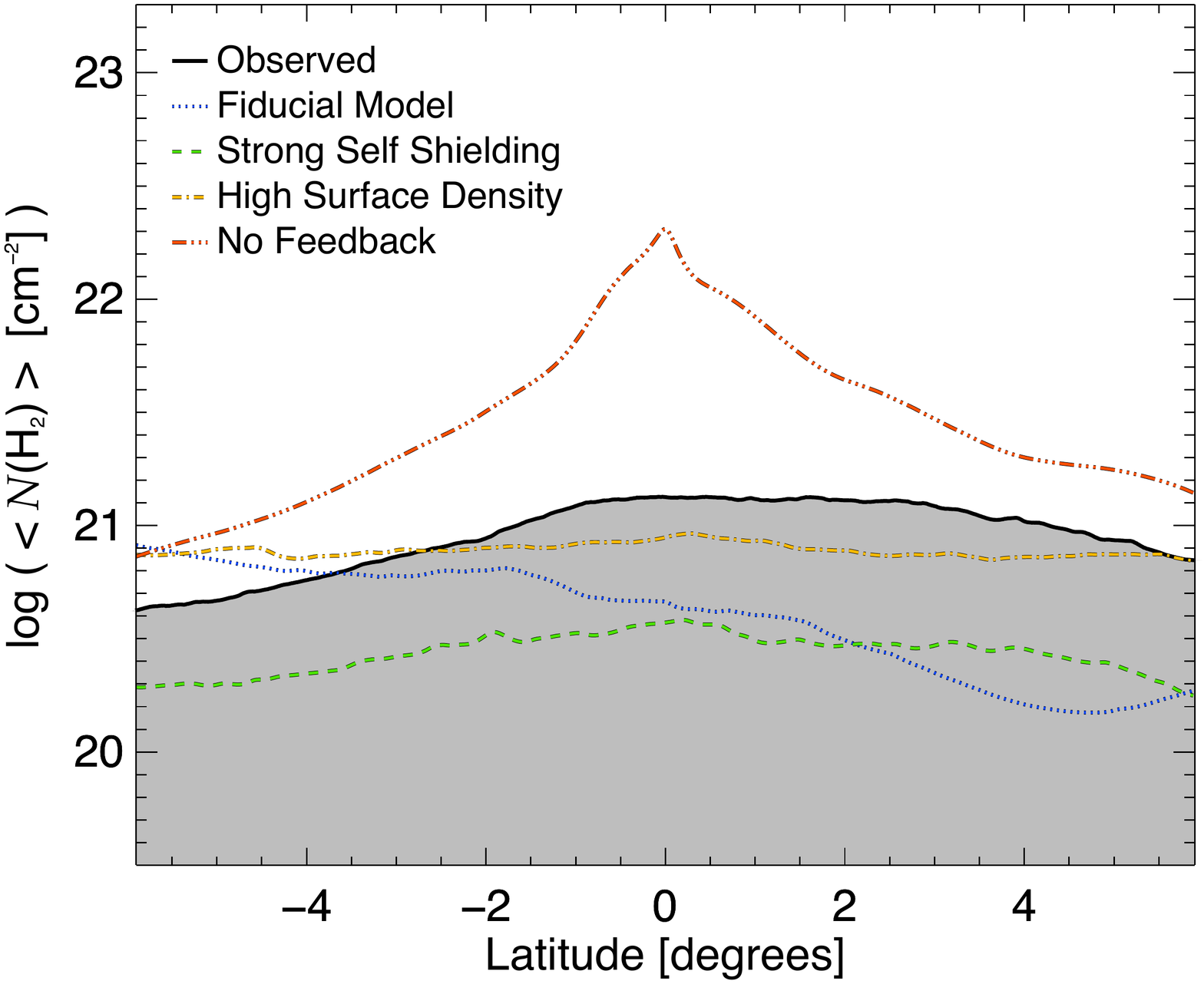}
	\hspace{-0.75cm}
	\includegraphics[width=0.36\textwidth]{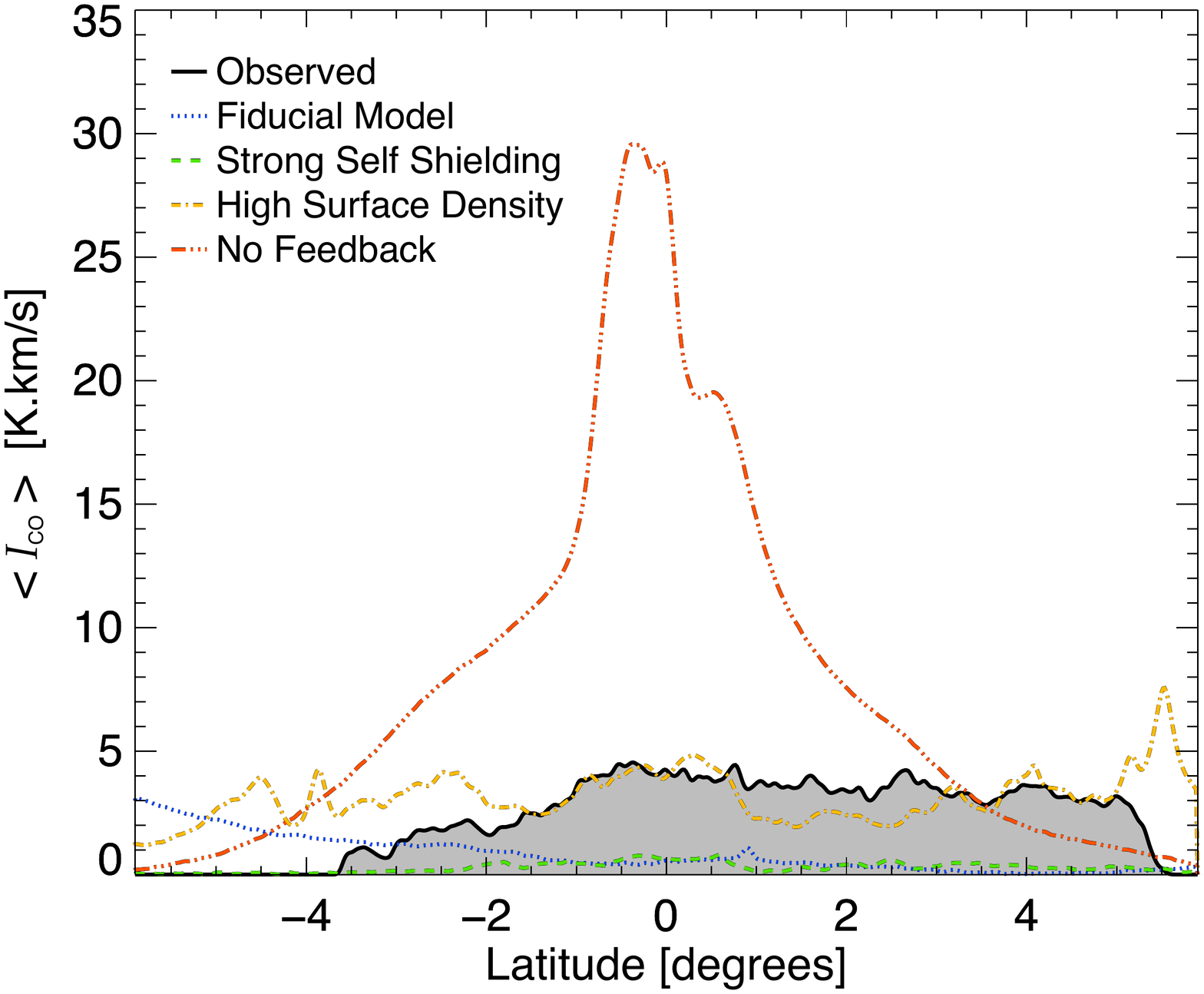}
\caption{Average latitude distribution of the H{\sc{i}} emission (left), the H$_{2}$ column densities (centre) and the CO intensities (right), for 2Q. The observations are shown in black solid lines and grey shaded areas, and the models are in coloured lines: fiducial in dotted blue line, strong self shielding in dashed green line, high surface density in dash-dotted yellow line, and the no feedback model in dash-triple-dotted red line. }
\label{fig:lat_dist}
\end{figure*}

Figures~\ref{fig:spatial_coldens} and \ref{fig:spatial_CO} show, for the four galaxy models we selected, the same as in Fig.~\ref{fig:fiducial_spatial}, i.e. the spatial distribution of the gas in the second galactic quadrant compared to the respective observed distributions in the Milky Way. While the runs with feedback produce plausible observed properties of H$_{2}$, H{\sc{i}} and CO, the model with no feedback fails to reproduce a realistic column density range and morphology. Moreover it fails to reproduce the full latitude extent of the emission, as the gas is collapsed onto the Galactic plane.

Figure~\ref{fig:lat_dist} shows this more clearly, with the comparison of the averaged latitude distribution of the emission for the four models in question, and the observations. We can see that the shapes of the distributions are similar for the observations and models, with the exception of the no feedback run, which is systematically more peaked around the galactic plane. This had already been noted by \citet{Acreman12}, based on H{\sc{i}} emission alone. The no feedback model also seems to significantly overproduce all three tracers. For the runs with feedback, the range of H$_{2}$ column densities and CO intensities are typically lower than those observed, while the synthetic H{\sc{i}} is systematically stronger. The high surface density model is the model that matches best the strength and amount of observed H$_{2}$ and CO emission, but it still overproduces the amount of atomic hydrogen (see last column of Table~\ref{tab:frac_molecular}), and consequently also overestimates the strength of the H{\sc{i}} emission.

\subsection{Velocity structure}
\label{section:lv_distributions}

\begin{figure*}
	\hspace{-0.4cm}
	\includegraphics[width=1.01\textwidth]{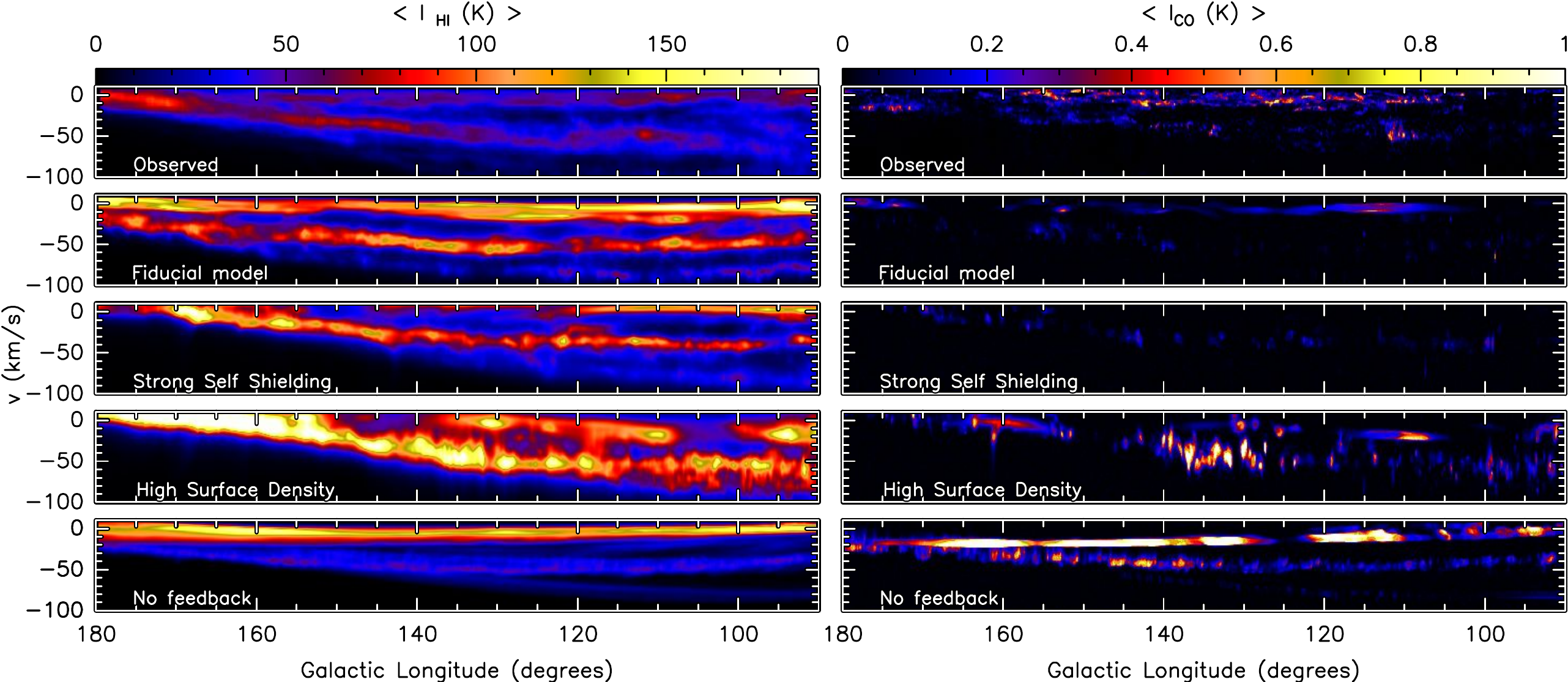} \\
  	\caption{Comparison of the observed l-v distribution of H{\sc{i}} ({\it left}) and CO ({\it right}) average intensity from the second quadrant of the Galaxy (top row) and the galactic models (on rows below). }
	\label{fig:velocity_HI}
	\label{fig:velocity_CO}
\end{figure*}

Figure~\ref{fig:velocity_CO} shows the comparison of the observed longitude-velocity averaged intensities of H{\sc{i}} and CO (top row), with the equivalent emission from the four simulations for which we have non-LTE CO calculations (panels below). Similarly to what we saw with the fiducial model (Sect.~\ref{section:fiducial_lv_distribution}), the galaxy model with a strong self shielding is able to trace the Perseus arm in a similar way to the observations, in a longitude-velocity space. However, it does not recover much local emission due to the specific location of the observer in this particular galaxy, which is just on the outer edge of its ``local'' arm (which can still be seen in H{\sc{i}}). This results in less material in the latitude-longitude distributions (Sect.~\ref{section:lb_distributions}), and we need to bear this in mind when comparing with observations. 

The high surface density model, however, shows more substantial differences. The main morphological difference is that the emission seems to be more concentrated in small bright patches, which reflects the fact that this model has formed more massive clouds. The larger amount of material and the existence of self-gravity makes the simulation more prone to form structures and hold them together, even whilst in inter-arm regions, resulting in a decreased contrast between arm and inter-arm regions (see right panel of Fig.~\ref{fig:top-down}), and consequently less defined arm structures in the respective longitude-velocity plot (see Fig.~\ref{fig:velocity_HI}, right column, fourth row). 

In the no feedback model, the arm/inter-arm regions have a sharper contrast, precisely because there is no self-gravity, and hence the spiral arm potential is the sole dictator of the distribution of material. The relative strength of the emission and its `smoother' continuous morphology, however, are not a very good match with the observations.

\subsection{CO and H$_2$}
\label{section:column_densities}

\begin{figure*}
  \hspace{-0.36cm}  	
   \begin{subfigure}[Fiducial Model]{
  	\includegraphics[height=3.8cm]{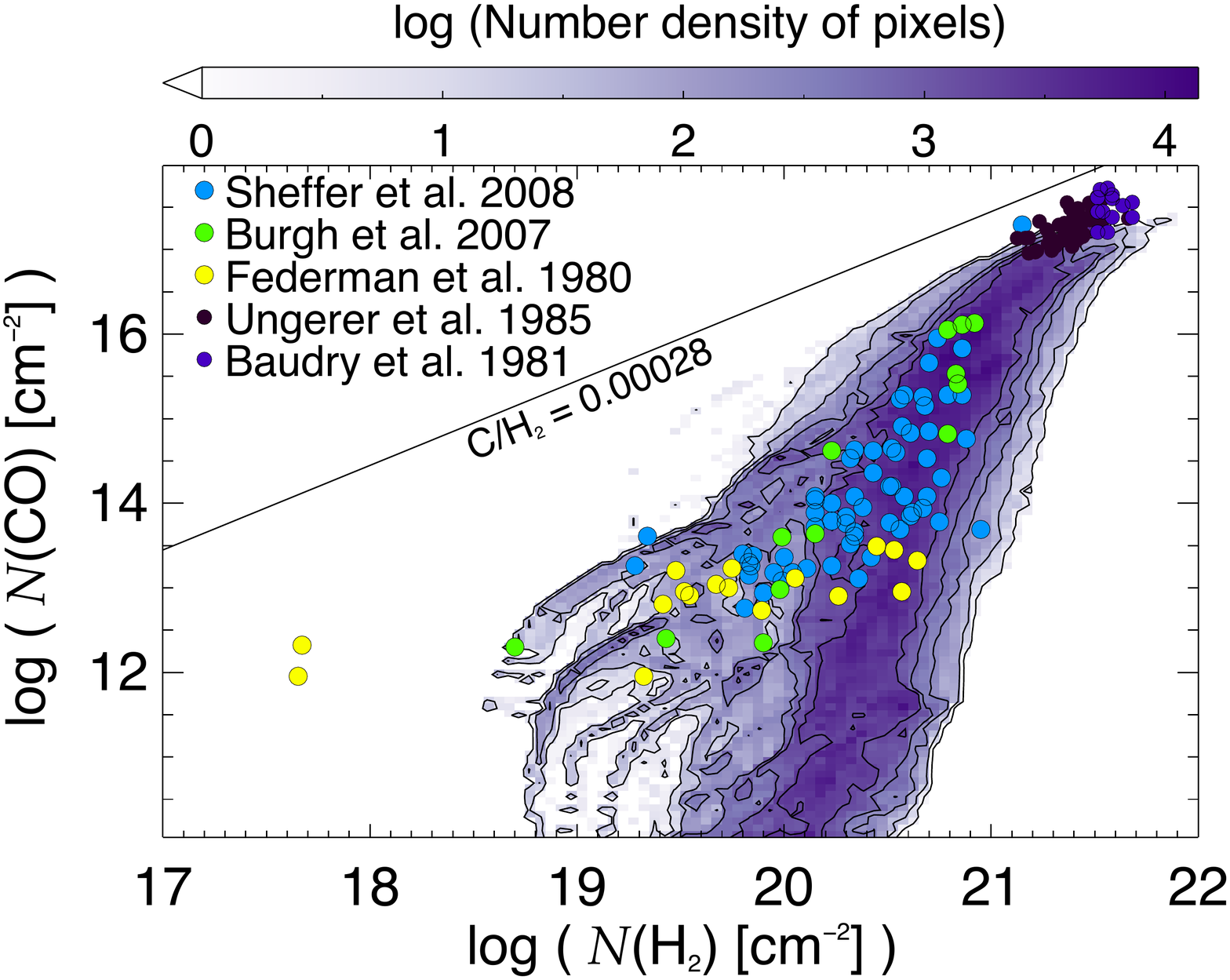}}
  \end{subfigure}
  \hspace{-0.5 cm}
  \begin{subfigure}[Strong Self Shielding]{
	  \includegraphics[height=3.8cm]{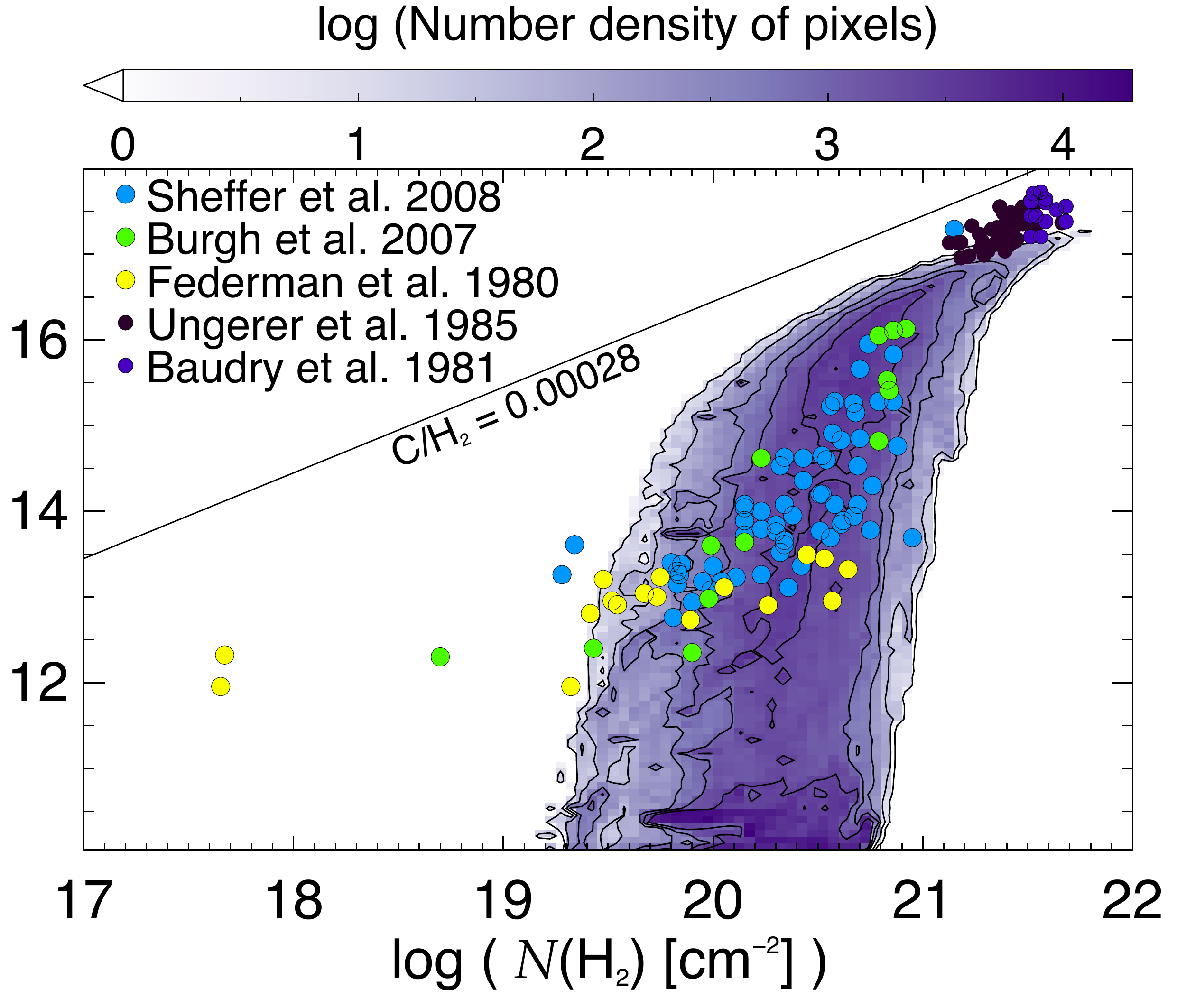}}
  \end{subfigure}
  \hspace{-0.5 cm}
   \begin{subfigure}[High Surface Density]{
  	\includegraphics[height=3.8cm]{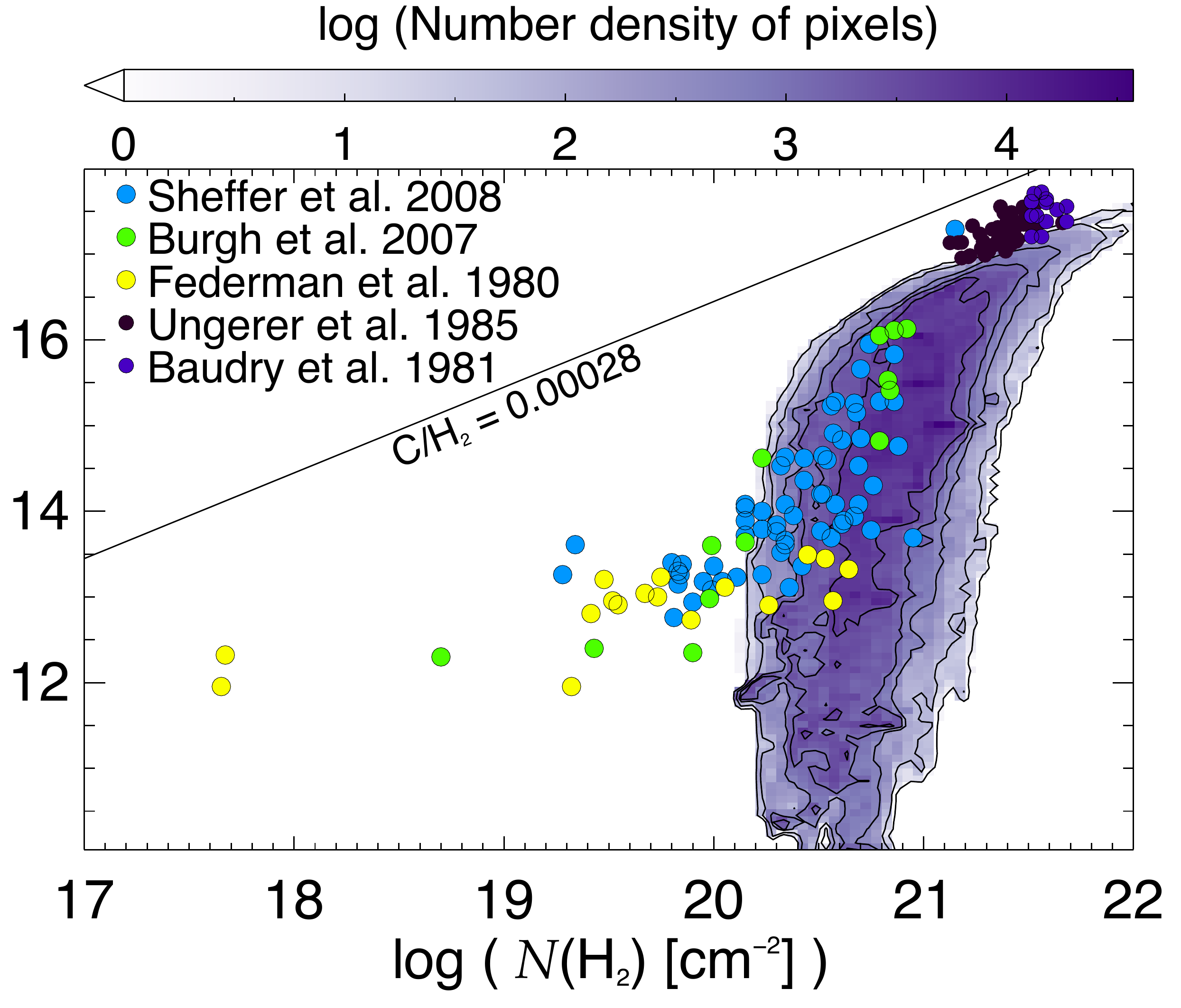}}
  \end{subfigure}
  \hspace{-0.5cm}
   \begin{subfigure}[No feedback]{  
   	\includegraphics[height=3.8cm]{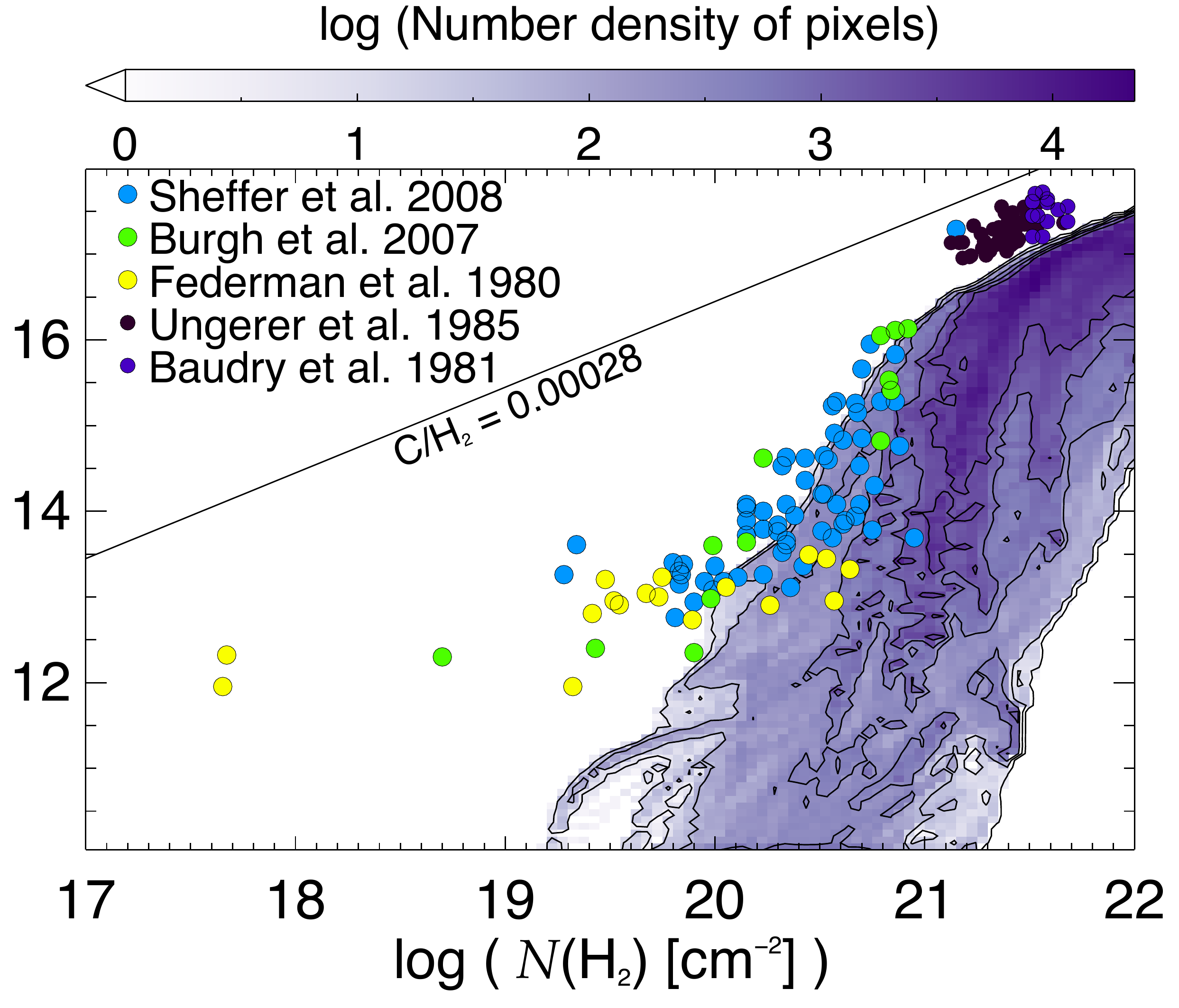}}
   \end{subfigure}
\caption{Distribution of the CO column density against the $\rm{H}_2$ column density from Galactic observations from the literature (coloured circles) and from the galaxy models (contours and colour scale).}
\label{fig:fuse}
\end{figure*}

As in Sect.~\ref{section:fiducial_column_densities}, we now compare the range of column densities from our four models to the observational determinations of Galactic molecular column densities in $\rm{H}_2$ and CO (Fig.~\ref{fig:fuse}). While the two lower surface density models with feedback (i.e. the fiducial and the strong self shielding models) both reproduce reasonably well the observed range of column densities, the higher surface density model seems to slightly over-predict the H$_{2}$ column densities with respect to the CO column densities. 
The no feedback run is the worst match between H$_{2}$ and CO, but this is easily understandable as: 1) the non-existence of feedback results in a higher concentration of material in the plane (as already seen in Sect.~\ref{section:lb_distributions}), which produces the high H$_{2}$ column densities for lines of sight along the galactic plane (which do not correspond to high volume densities); 2) the non-existence of self-gravity yields a production of CO much less efficient than it would be if self-gravitating clouds are able to form, and effectively reach higher volume densities.

\begin{figure*}
 \hspace{-0.38cm}
  \begin{subfigure}[Fiducial model]{
  	\includegraphics[height=4.5cm]{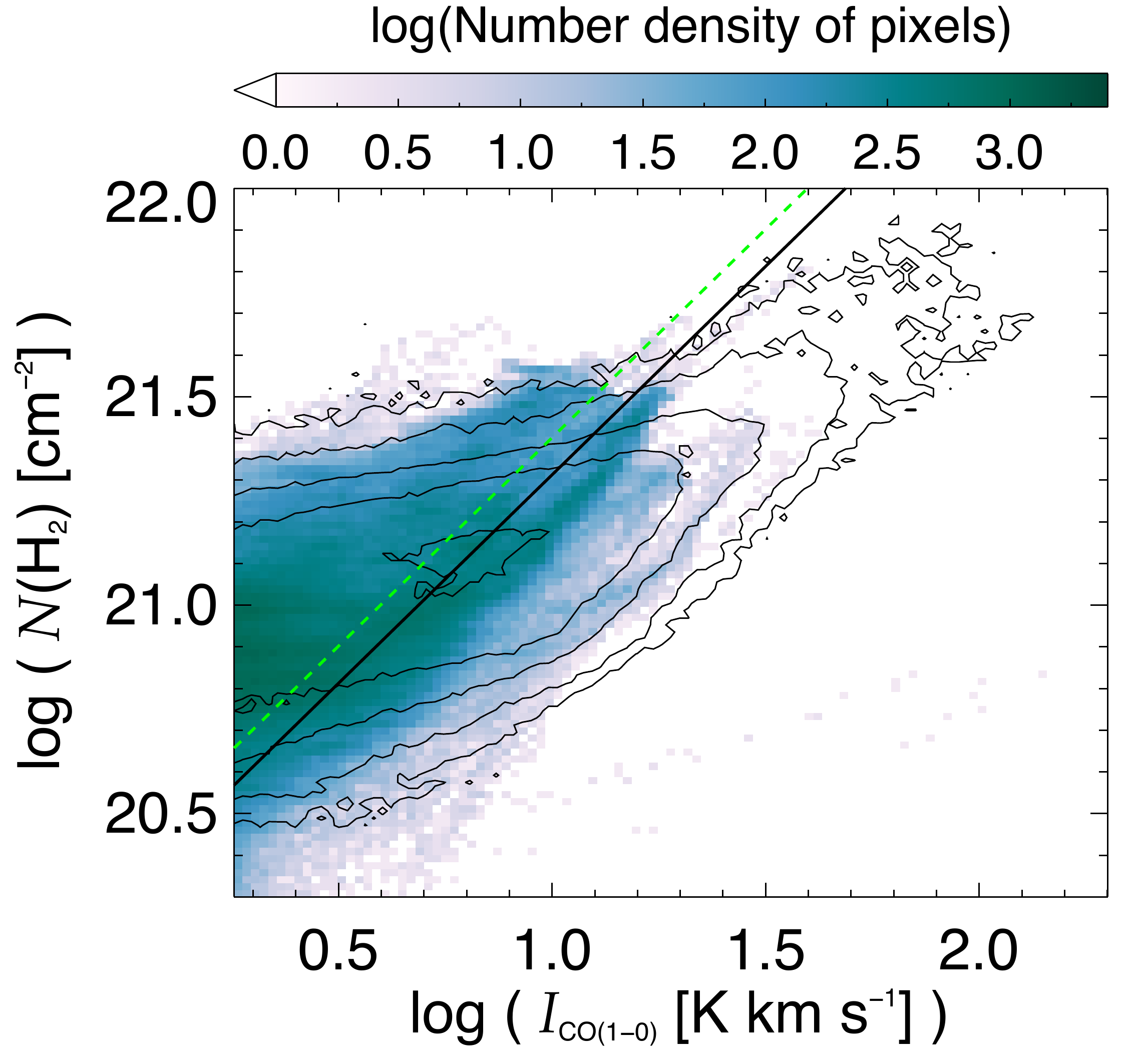}}
  \end{subfigure}
   \hspace{-0.5cm}
   \begin{subfigure}[Strong Self Shielding]{
	 \includegraphics[height=4.5cm]{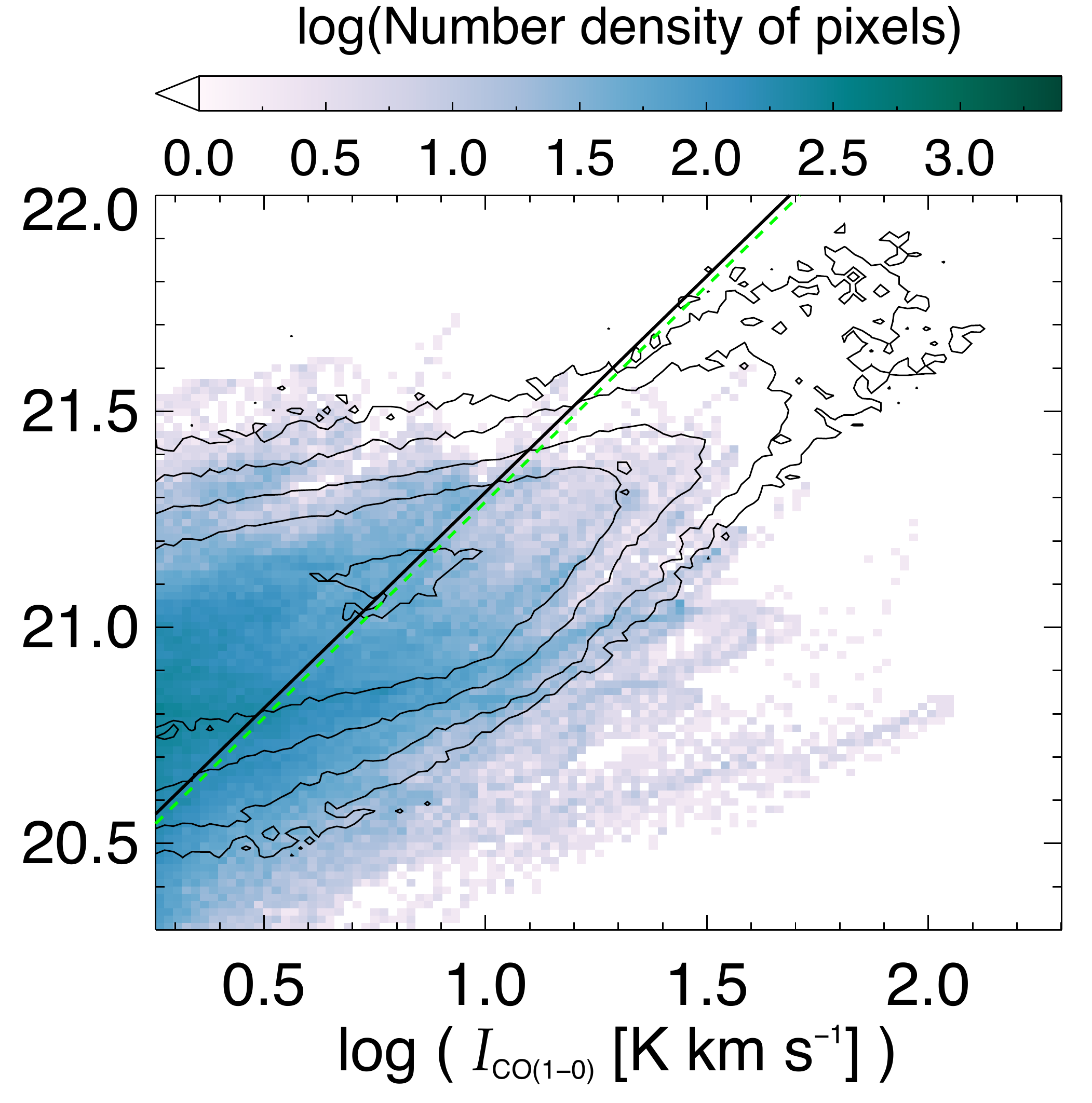}}
  \end{subfigure}
     \hspace{-0.5cm}
  \begin{subfigure}[High Surface Density]{
  	\includegraphics[height=4.5cm]{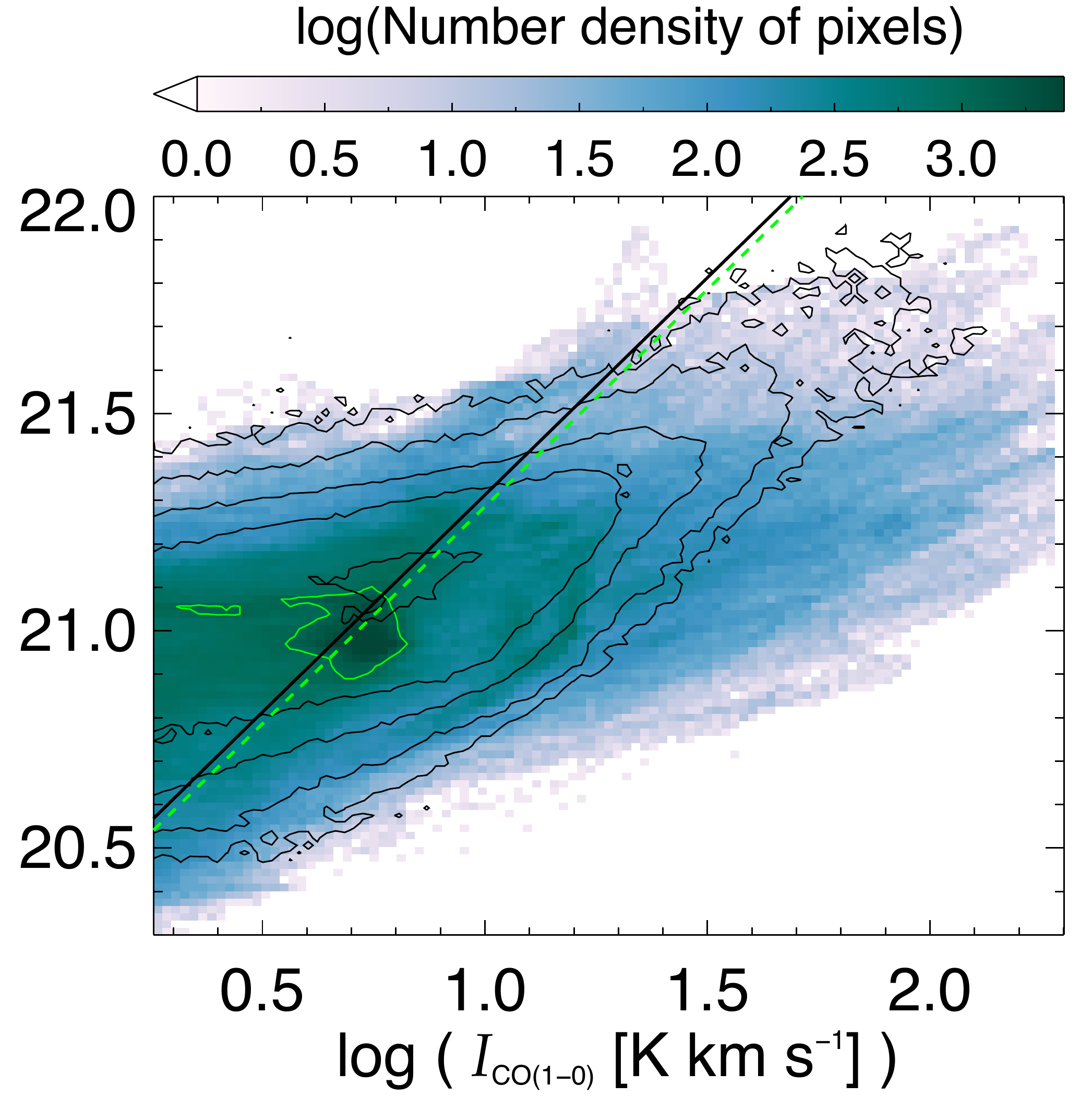}}
  \end{subfigure}
     \hspace{-0.5cm}
   \begin{subfigure}[No Feedback]{  
   	\includegraphics[height=4.5cm]{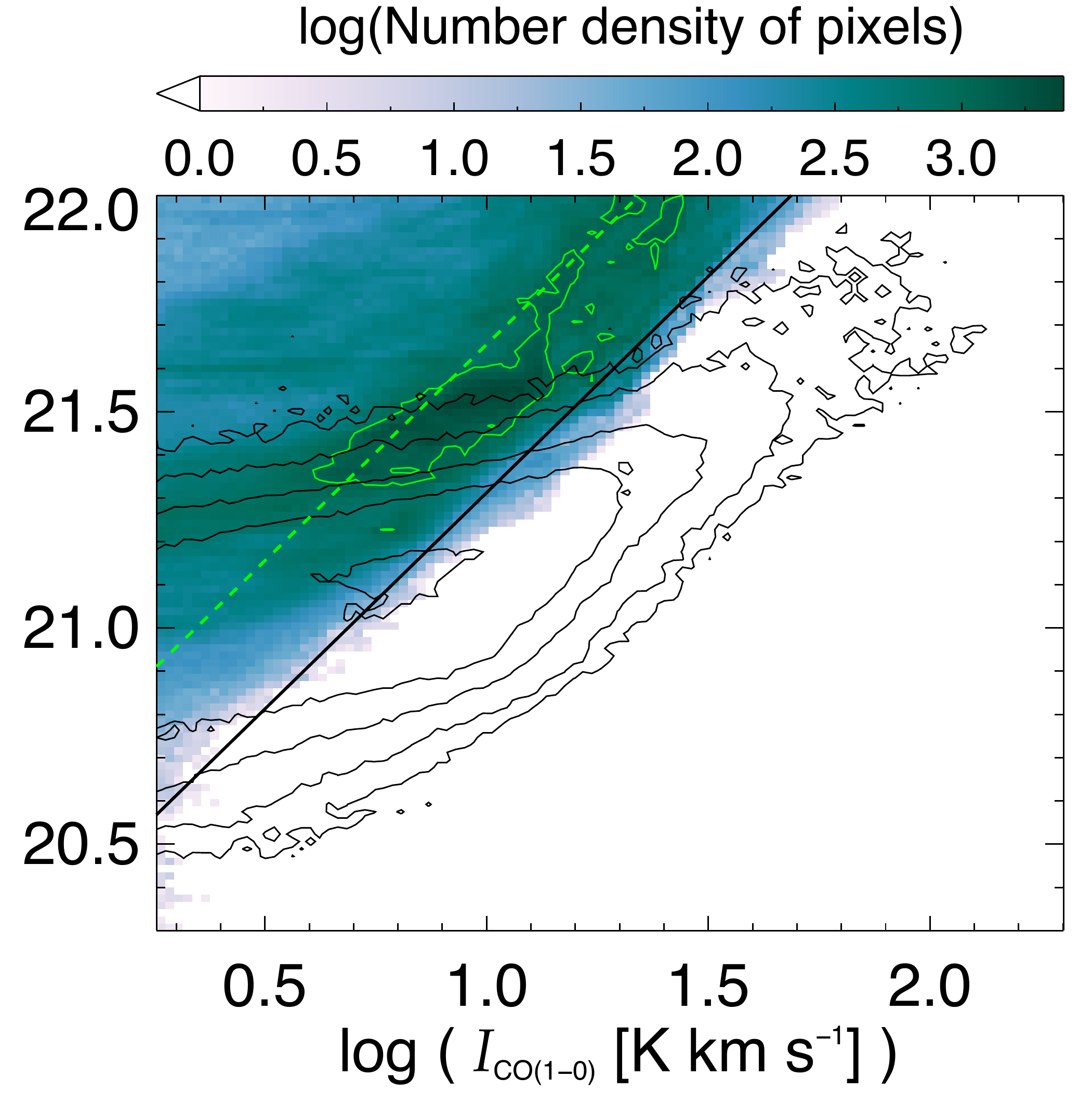}}
   \end{subfigure}
\caption{Distribution of the CO intensities against the $\rm{H}_2$ column density from the observations (in contours) and the four models (colour scale). The black dashed lines show the median X$_{\rm CO}$ factor from the observations, and the red dashed lines correspond to the median X$_{\rm CO}$ for each model (see Table~\ref{tab:frac_molecular}).}
\label{fig:CO_nH2}
\end{figure*}

In terms of the correspondence between CO intensities and $N(\rm H_{2})$ column densities (see Fig.~\ref{fig:CO_nH2}, and X$_{\rm CO}$ values in the second column of Table~\ref{tab:frac_molecular}, estimated as described in Sect.~\ref{sec:Xfactor_fiducial}), we find that the X$_{\rm CO}$ values for the three models with feedback are all typically consistent with the best fit value for the observations \citep[and the value of $\sim 1.8 \times 10^{20}$\,cm$^{-2}$\,K$^{-1}$\,km$^{-1}$\,s estimated by][]{Dame01}. A comparison of the distributions in colour and contours in Fig.\ref{fig:CO_nH2} (i.e. the observed vs. model distributions) shows that the strong self-shielding and high surface density models appear to have a marginally better correspondence than the fiducial model. Again, it is clear that the run with no feedback is the one most offset from the observations, with a significantly higher X$_{\rm CO}$. 

This suggests that whenever feedback and self-gravity are present, then changes in surface density, or the feedback/chemistry parameters do not have a big impact on the way that H$_{2}$ is correlated with CO. Instead, the dominating effect on the efficiency of forming CO is the existence of both self-gravity (which would allow more CO to be formed in the denser clouds), and feedback (which distributes the emission with latitude, essential to reproduce the line of sight column density values retrieved in our Galaxy).

\begin{figure*}
  \hspace{-0.4cm}
  \begin{subfigure}[Fiducial model]{
  	\includegraphics[height=4.55cm]{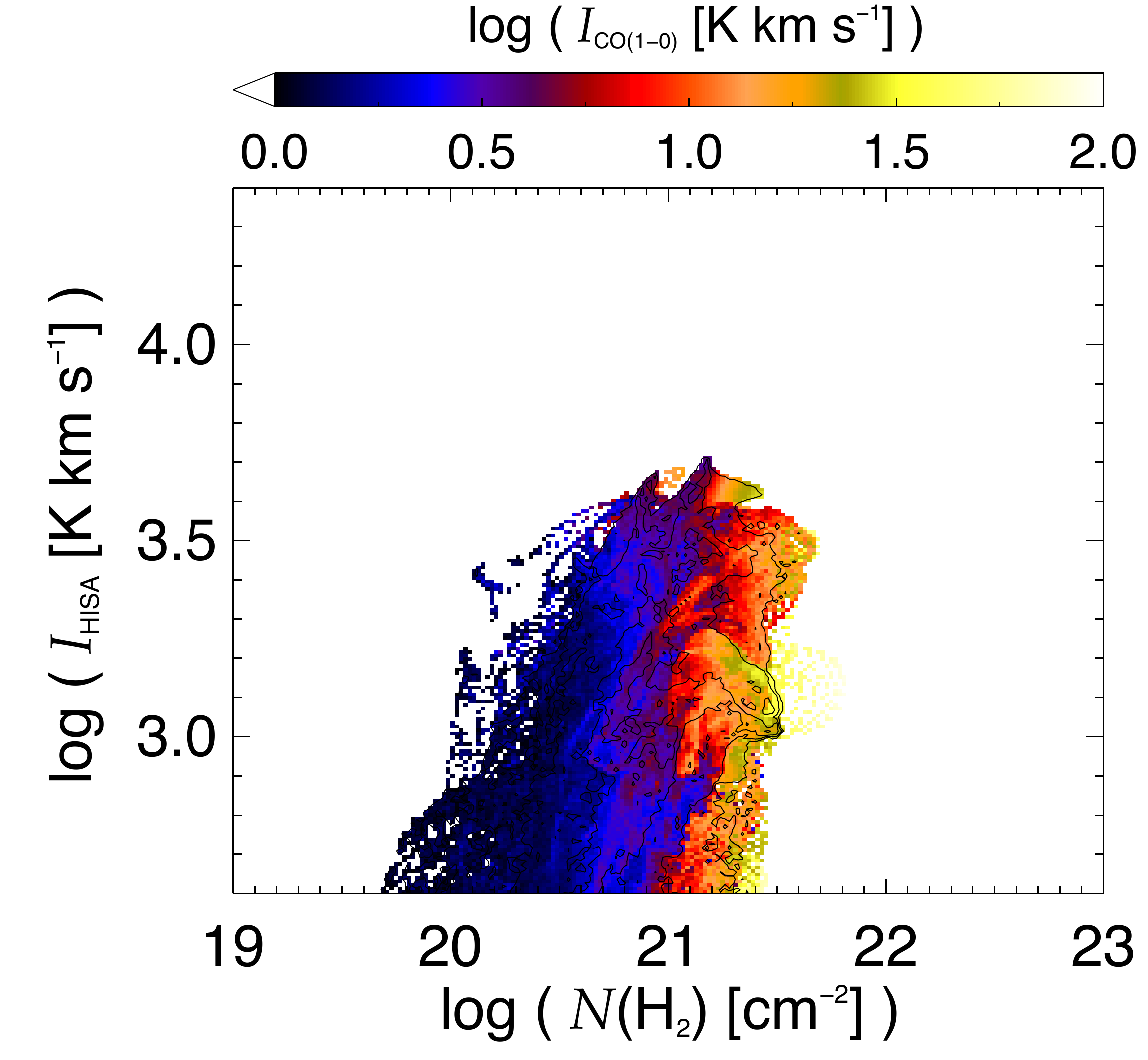}}
  \end{subfigure}
  \hspace{-0.55cm}
   \begin{subfigure}[Strong Self Shielding]{
	 \includegraphics[height=4.55cm]{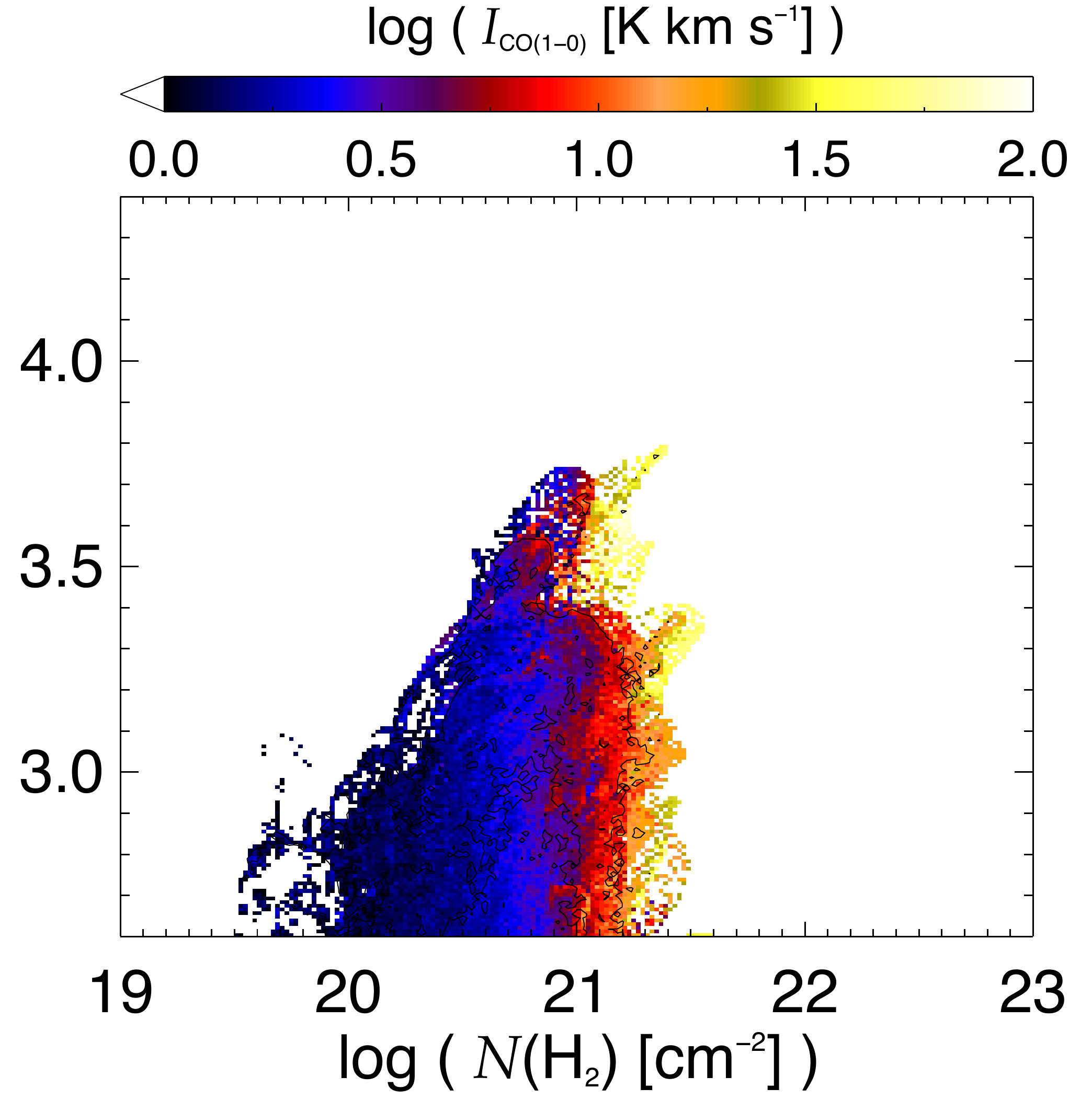}}
  \end{subfigure}
  \hspace{-0.55cm}
   \begin{subfigure}[High Surface Density]{
	 \includegraphics[height=4.55cm]{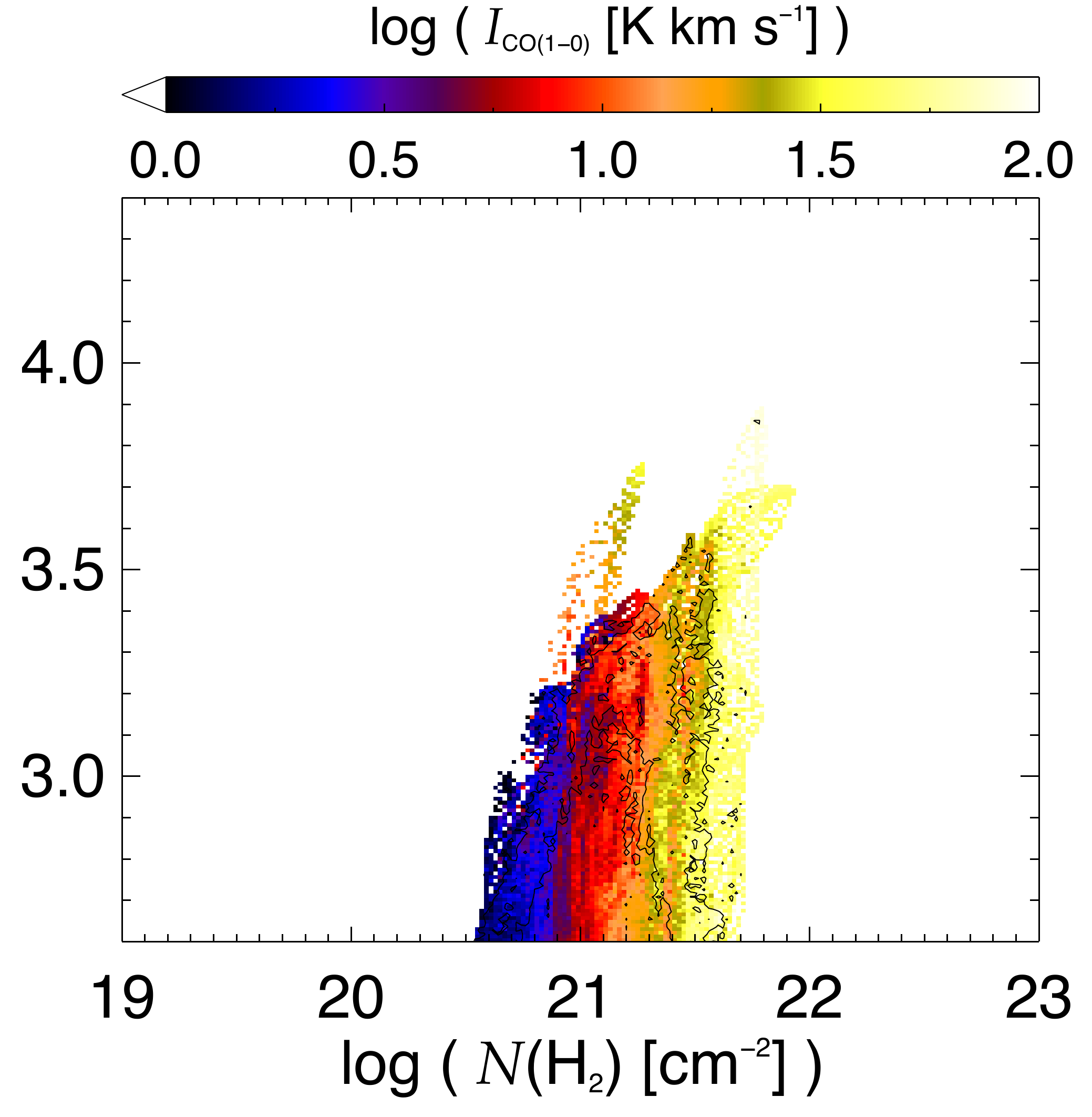}}
  \end{subfigure}
  \hspace{-0.55cm}
   \begin{subfigure}[No feedback]{  
   	\includegraphics[height=4.55cm]{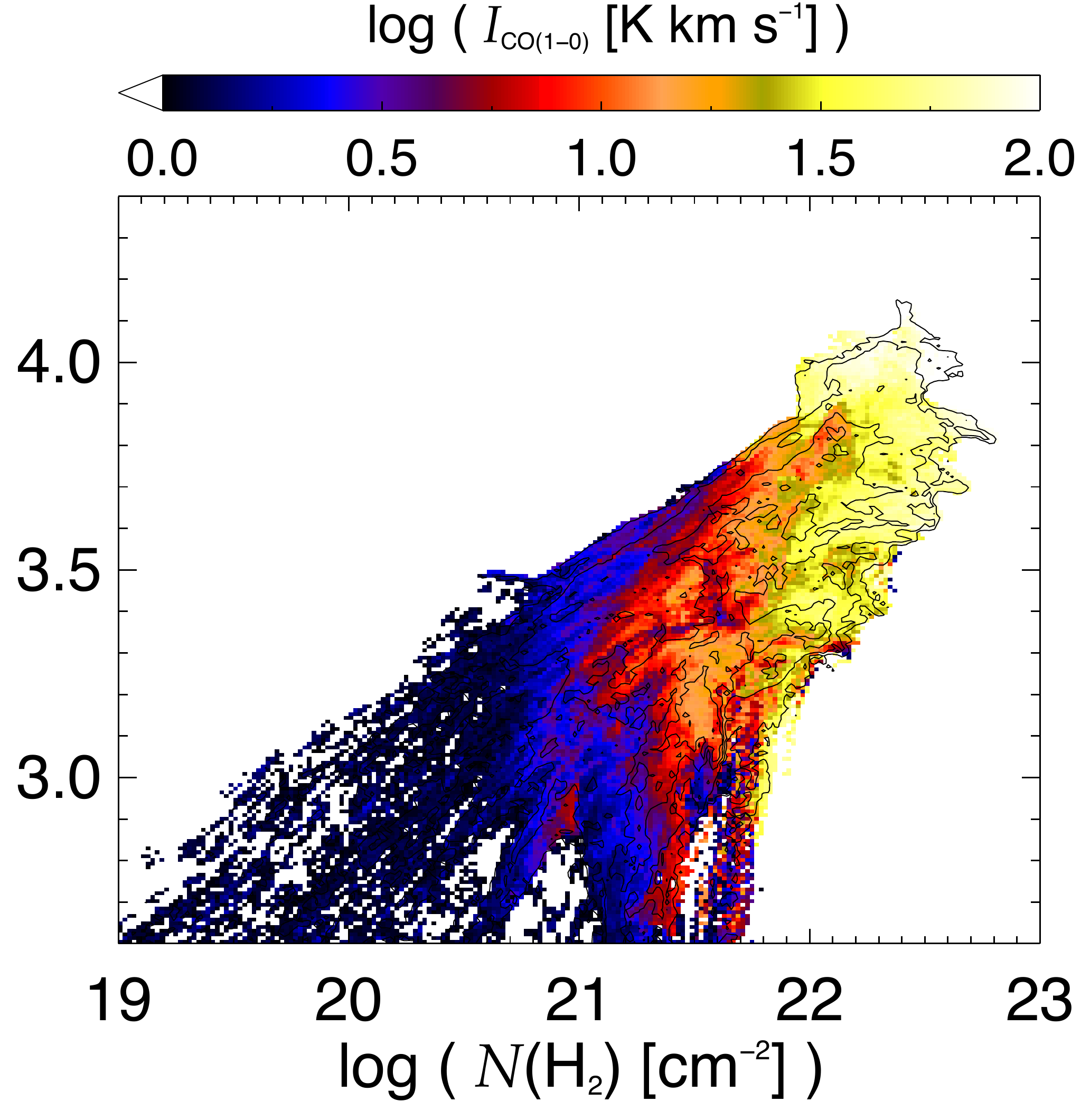}}
   \end{subfigure}
\caption{Distribution of the absolute value of the HISA integrated intensities, against the corresponding $\rm{H}_2$ column densities, colour-coded with the CO integrated intensities, for the four galaxy models. Contours follow the density of pixels for each $I_{\rm HISA}$-$N(\rm{H}_2)$ bin.}
\label{fig:HISA_nH2_CO}
\end{figure*}

\subsection{HISA, CO and $\rm{H}_2$}
\label{sec:HISA_CO_H2}

\begin{table*}
\caption{Statistical properties from the observations and the four galactic simulations.}
\label{tab:frac_molecular}
\flushleft
\renewcommand{\footnoterule}{}  
\begin{tabular}{l c c c c l c c }
\hline \hline	
\multirow{2}{*}{Name	}	& X$_{\rm CO}$					&  	$\alpha$  				& $M($H$_{2})^{\rm CO}$	& $M($H$_{2})^{\rm SPH}$$^{(a)}$ & Fraction$^{(b)}$ &  H/H$_{2}$$^{(a)}$\\
					& [$10^{20}$ cm$^{-2}$\,K$^{-1}$\,km$^{-1}$\,s] 	& (from $M\propto S^{\alpha}$)	& [$10^5$ M$_{\odot}$]	& [$10^5$ M$_{\odot}$]	   & & \\
\hline
Observations			& $2.0\,(\pm\,0.9)$	& $1.1\pm0.1$	&	$33$			&	-	 	&	$20-80\%$$^{(c)}$	& $\sim 6^{(d)}$ \\	
Fiducial		 		& $2.5\,(\pm\,0.9)$	& $1.8\pm0.6$	&	$5$			&	10 		&	$\sim 50\%$			& 85 \\	
Strong Self Shielding 	& $1.9\,(\pm\,0.9)$	& $2.4\pm0.5$	& 	$5$			&	13	 	&	$\sim 40\%$			& 80 \\
High Surface Density 	& $1.9\,(\pm\,0.8)$	& $2.2\pm0.3$	&	$139$		&	60 		&	$\sim 100\%$$^{(e)}$	& 29 \\
No Feedback			& $4.5\,(\pm\,1.9)$	& $1.6\pm0.3$	&	$74$			&	235	 	&	$\sim 30\%$			& 6 \\
\hline
\end{tabular}
\flushleft
$^{(a)}$ Ratio of atomic to molecular hydrogen, estimated from the SPH output, for the 2$^{\rm nd}$ Quadrant below 160$^{\circ}$ longitude.\\
$^{(b)}$ Percentage of molecular material traced by CO, i.e., the ratio of $M($H$_{2})^{\rm CO}\,/\,M($H$_{2})^{\rm SPH}$.\\
$^{(c)}$ From \citet[][]{Pineda2013}, for the entire Galactic disk.\\
$^{(d)}$ From estimates of the total mass of atomic hydrogen in our Galaxy \citep[by][]{Wolfire03}, and the total mass of H$_{2}$ \citep[e.g. from][]{Williams1997,Bronfman2000}. \\
$^{(e)}$ The high surface density model was the model that suffered the most severe over-estimations of distances, which ultimately resulted on the overestimated molecular mass from CO, off by more than a factor 2 from the total molecular mass existent in the simulation. 
\end{table*}

Following what we saw in Sect.~\ref{sec:HISA_CO_H2_fiducial}, we now investigate the statistical spatial relationship between HISA, CO and $N(\rm{H}_2)$, for the four models. Figure~\ref{fig:HISA_nH2_CO} shows the scatter plot of the absolute value of the HISA integrated intensities, against the corresponding $\rm{H}_2$ column densities, colour-coded with the CO integrated intensities.  From this figure we see the tentative correlation between $I_{\rm HISA}$ and the $N(\rm{H}_2)$ from the models, even though the dispersion is quite large (approximately an order of magnitude). For all the models with feedback, the CO intensities seem completely uncorrelated with the HISA intensities, instead depending solely on the column density of H$_{2}$. CO starts appearing at column densities above $\sim 10^{21}$\,cm$^{-2}$, and whenever CO starts to be detected, it does not show any gradient of intensities along the y-axis (i.e. with $I_{\rm HISA}$), and instead, it varies with the x-axis (i.e., with $N(\rm{H}_2)$). The three galaxy models with feedback have similar distributions. Once again, the run with no feedback stands out and appears to have a tentative correlation between HISA and $N(\rm{H}_2)$. Because of the higher concentration of material (both atomic and molecular) along the galactic plane, this model naturally presents better conditions for stronger HISA, and therefore it reaches much higher $I_{\rm HISA}$ than observed. 
However, since this no feedback model is the least realistic model we are testing, it is hard to judge whether this correlation is meaningful.

\subsection{Molecular gas traced by CO}
\label{sec:molecular_by_CO}

To investigate how well we can recover the total amount of molecular gas by using CO as a molecular gas tracer, we have compared the total mass of molecular gas by summing the mass of all the CO clouds extracted for each of the galaxy models as detailed in Sect.~\ref{sec:cloud_properties_fiducial}, and compared that to the actual molecular gas mass contained in the second quadrant.  The results from this exercise are summarised in Table~\ref{tab:frac_molecular}.

We are able to recover between 30 and 100$\%$ of the total molecular mass existent in the simulations. The run with strong self shielding produced similar results to the fiducial model, although the different resolutions may contribute to the slight differences, as lower mass clouds are harder to detect in the strong self shielding model. From the observations of  \citet[][]{Pineda2013} it was estimated that in the Milky Way the fraction of molecular gas which is efficiently traced by CO can vary between 20$\%$ and 80$\%$. That study suggests that the fraction of molecular gas not traced by CO increases towards the outer regions of the Milky Way. Therefore, since when observing the second galactic quadrant we are looking at the outer parts of the Galaxy, it is perhaps not surprising to see low fractions of molecular gas being traced by CO. The only simulation that has a higher fraction is the high-surface density run which recovers all the molecular content through the CO clouds. This high fraction may be an artefact, since the extracted clouds are `biased' towards higher masses (due to the SPH resolution), while the X$_{\rm{CO}}$ used for converting from CO to H$_{2}$ was calculated using the entire column density spectrum of H$_{2}$, which is dominated by intermediate column densities of $10^{21}$\,cm$^{-2}$. For higher column densities the X$_{\rm{CO}}$ shows a tendency to decrease, and therefore, the total molecular mass in these high-density clouds may be overestimated when using CO in this manner. 

This table also shows the estimated values of ${\alpha}$ from fitting the mass-size relation ($M \propto S^{\alpha}$) for each dataset, as in Sect.~\ref{sec:cloud_properties_fiducial}. We caution, however, that the worse mass resolution for the strong self shielding and high surface density models could potentially alter the distribution of masses and sizes, perhaps leading to the observed higher ${\alpha}$ values. Nevertheless, we find that despite the different ${\alpha}$, the peak of the mass and size distributions of the two low surface density runs with feedback (fiducial and strong self shielding) are similar to the observed distributions ($\sim 10^{3}$\,M$_{\odot}$ and $\sim20$\,pc$^{2}$). The other two runs, however, peak at higher masses ($\sim 10^{4}$\,M$_{\odot}$) and larger sizes ($\sim100$\,pc$^{2}$) compared to the observations. While for the high surface density run this is most likely simply due to the fact that we cannot resolve the lower mass clouds, this is not the case for the no feedback run, as it has the same resolution as the fiducial model. Instead, this shift is a consequence of the absence of feedback to break up the material and create smaller substructures \citep[in line with what had been reported in][]{Dobbs11b}.

\subsection{Other simulations}
\label{sec:other_sims}

As mentioned in Sect.~\ref{section:galaxy_model}, we also ran a few other simulations which we do not show here, to test other free parameters of the simulations. 

First we ran a calculation where we did not use the mass of molecular gas to insert feedback, rather we used the total mass of gas in neighbouring particles. This was to test whether we were reducing the amount of CO by preferentially inserting feedback into molecular regions. Further, when we insert feedback, we force-set the abundances of H$_{2}$ and CO in those particles to low values, assuming that feedback automatically destroys those species. We performed one test where we did not switch the abundances to low values. Finally, we also analysed an earlier time-step of this later model, as well as of the model with high surface density discussed above. The time-step chosen is when the amount of H$_{2}$ and CO reaches a maximum, which corresponds to the moment where feedback has not yet started, and hence these are equivalent to simulations without feedback but with self-gravity included. 

We did not run non-LTE calculations for the CO emission, but we have compared the H$_{2}$, CO and H{\sc{i}} distribution for these four extra models. With respect to the fiducial model, we found very little variation for the two extra models with feedback. The latitude extent, as well as the amount of CO and H$_{2}$ formed are very similar to what we saw for the fiducial and strong self shielding models, which means that our results are not very sensitive to changes in the recipes for feedback implementation. On the other hand, the two models with self-gravity but no/little feedback, are rather different. In fact, their behaviour and distributions are similar to the model with no feedback that we have studied in this paper, with the exception of having the emission more `blobby' as a consequence of self-gravity holding the gas together, and the non-existence of a mechanism to break it apart. In particular, the high surface density model before feedback looses nearly all contrast between arms and inter arm regions, as self-gravity on individual cloud-scales dominates over the gravitational potential of the galaxy. This accentuates the fact that both feedback and self-gravity are essential to implement in (galactic) simulations, in order to be able to reproduce the basic chemical properties of the ISM in spiral galaxies.

To test whether the density threshold imposed by the implementation of feedback could be limiting the formation of H$_{2}$ from atomic H, we ran a final test where we increased this density limit for the insertion of feedback, and reran the last 50\,Myr of a simulation. Doing this does increase the amount of H$_{2}$ with respect to atomic H, supporting the idea that the high H/H$_{2}$ ratio in the current models is partly a consequence of the numerical limitations, rather than an intrinsic problem in the way atomic H is converted into H$_{2}$. 

\section{Summary and conclusions}
\label{section:conclusions}

In this article we have studied the current ability to reproduce the basic ISM large scale structure and chemical composition, by comparing the emission of a model of a grand design galaxy to observations of our own Milky Way. We found that the output from our fiducial model agrees well with the observations, obtaining similar values for X$_{\rm CO}$, and similar distributions of emission, which means that the current models provide a good approximation for the formation of CO from H$_{2}$. Due to numerical limitations, however, the model fails to reproduce the detailed morphology of the CO emission (it is more compact and sparse than in observations), and it is unable to resolve the lower density filamentary structures that we see in the Galaxy. It also fails to reproduce the right ratio between molecular and atomic hydrogen (underproducing H$_{2}$ from atomic H on a galactic scale). 

We also tested how sensitive our results were to the feedback and chemistry recipes. For this, we have compared a series of models against the fiducial model. This comparison essentially shows that the variations that we applied on the feedback and/or chemistry recipes do not change the output hugely. However, the implementation of feedback is essential as otherwise all the chemical properties are clearly offset from observations. 

We also tested how increasing the mass of the simulated galaxy could impact the ability to efficiently form H$_{2}$ from atomic H, and we found that an increased mass decreases the ratio of H/H$_{2}$ (as it becomes easier to form larger and denser clouds and, therefore, H$_{2}$ is more efficiently formed), but it is still rather high compared to the observations. The reasons for this problem are not yet clear, and are yet to be elucidated. This is likely a consequence of both the resolution and the density (and temperature) thresholds applied. 
The lack of resolution likely leads to an overestimate of the volume influenced by the feedback events, therefore overestimating the amount of molecular material which is destroyed. Furthermore, setting a relatively low density threshold for the implementation of feedback will limit the production of molecular material, as we are never allowing the gas to become as cold and as dense as the typical molecular clouds observed in our Galaxy. This could imply that simulations will have difficulty correctly following both CO/H$_{2}$ and stellar feedback as they will not be able to resolve the total CO content, particularly in a cosmological context. These effects will be investigated in follow up papers. 

Overall, we conclude that the surface density and presence of feedback and gravity have a large effect on the basic chemical properties of the ISM, while changing the chemistry/feedback recipes has a minimal impact.

\section*{Acknowledgments}

We thank the referee, Ralf Klessen, for his comments that helped strengthen the paper. ADC and CLD acknowledge funding from the European Research Council for the FP7 ERC starting grant project LOCALSTAR. The calculations for this paper were performed on the DiRAC Complexity machine, jointly funded by STFC and the Large Facilities Capital Fund of BIS, and the University of Exeter Supercomputer, a DiRAC Facility jointly funded by STFC, the Large Facilities Capital Fund of BIS and the University of Exeter. Figure~\ref{fig:top-down} was produced using SPLASH \citep{Price2007}. We also thank A. Rodrigues for providing high-resolution dust column density maps for benchmarking.

\bibliographystyle{mn2e}
\bibliography{fb} 

\appendix

\section{Separating the Local and Perseus arms for the Galactic 2Q}
\label{app:arms_obs}

As described in Sect.~\ref{sec:HISA_CO_H2_fiducial}, we separated the emission of CO, HISA and the H$_{2}$ column densities arising from each arm along the line of sight, so as to avoid comparing the emission from material of different but overlapping structures. The same method was used for both observations and simulations. 

For CO emission and HISA, this separation was done by integrating the velocities respective to each arm (associating everything above -20\,km\,s$^{-1}$ to the Local arm, and everything below to the Perseus arm). For the H$_{2}$ column densities, since we have no velocity information to distinguish between the different arms, we used the CO emission at each pixel to evaluate which arm is likely contributing the most to the observed H$_{2}$ column densities. In practice, for any given pixel, we evaluated the arm with the stronger CO emission and compared that to the emission from the weaker arm. We associate all the H$_{2}$ column density to the stronger arm whenever its CO emission is a minimum of three times stronger than on the weaker arm, or whenever the emission of the weaker arm is below the CO noise level. If no strong CO component can be distinguished, we cannot separate the column densities properly, and therefore refrain from using such pixels.

\begin{figure*}
\flushleft
  	\includegraphics[width=\textwidth]{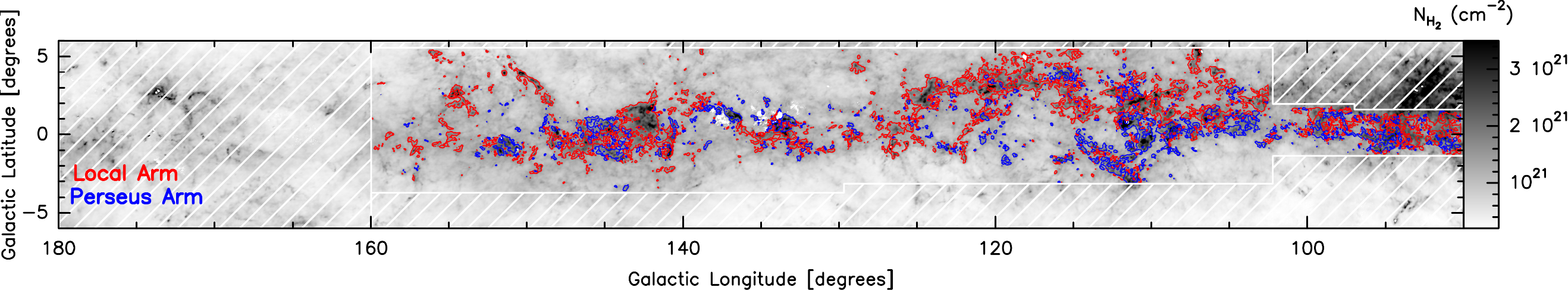} 
  	\caption{Spatial distribution of the observed Perseus (blue) and Local (red) arms in H$_{2}$ column densities (total H$_{2}$ column densities shown in grey scale). The white dashed areas represent regions where there is no CO coverage, plus data points above a longitude of 160$^{\circ}$ that were not used for constructing the scatter plot shown in Fig.~\ref{fig:HISA_nH2_CO_fiducial}, due to the degeneracy in velocities, which make the separation of the different arms unreliable.}
	\label{fig:local_perseus_ncol}
\end{figure*}

\begin{figure*}
\flushleft
  	\includegraphics[width=0.99\textwidth]{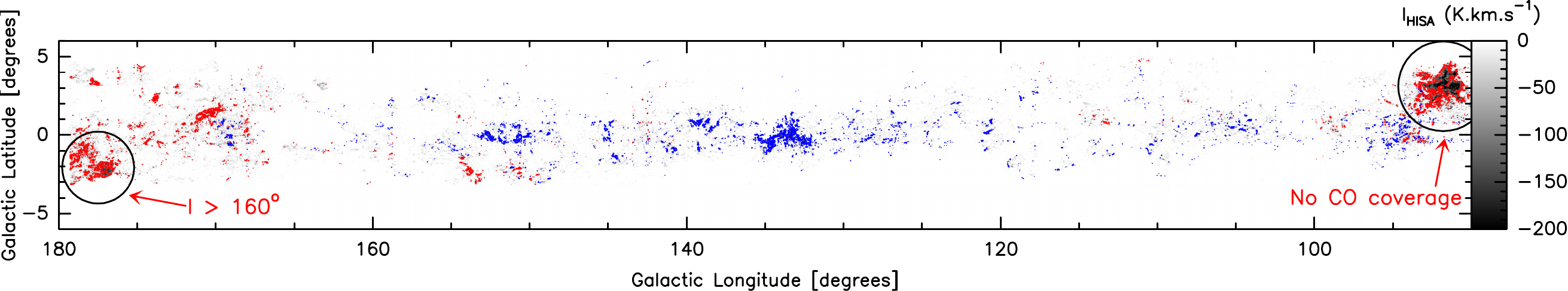} 
  	\caption{Spatial distribution of the observed Perseus (blue) and Local (red) arms in HISA intensities (total integrated HISA intensity is shown in grey scale).}
	\label{fig:local_perseus_hisa}
\end{figure*}

The resulting maps of the observed H$_{2}$ column densities and HISA emission can be seen in Figs.~\ref{fig:local_perseus_ncol} and \ref{fig:local_perseus_hisa} respectively, where the blue contours delineate regions selected as part of the Perseus arm, and the red contours delineate regions associated to the Local arm. Note that the area of the $N$(H$_{2}$) map is more restricted than the original, because it is constrained to the positions where we have CO emission, and to regions with longitudes below 160$^{\circ}$. 

From Fig.~\ref{fig:local_perseus_hisa}, we can see that the two strongest HISA features lie in regions that we did not use (see labeled black circles), either because they are not covered by the CO map, or because they lie at longitudes higher than 160$^{\circ}$. The only region that still has significant HISA and is covered by all maps (at $l\sim135^{\circ}$), actually includes some pixels that had been masked from the original SED fitting because of higher temperatures. Furthermore, it also includes some pixels where the separation from the two arms was not successful, as there was significant CO emission at both velocity ranges. All in all, this makes perhaps these observations of the second Galactic quadrant non ideal for testing the pixel-by-pixel correlation between HISA and $N$(H$_{2}$) or CO, and this could be part of the reason why there is no correlation on the observational distributions of Fig.~\ref{fig:HISA_nH2_CO_fiducial} (Sect.~\ref{sec:HISA_CO_H2_fiducial}).

\end{document}